\documentclass[onecolumn,aps,prd,preprintnumbers,showpacs,superscriptaddress,nofootinbib,amsmath,amssymb,floats,floatfix,showkeys,notitlepage,longbibliography]{revtex4-1}

\usepackage{orcidlink}
\usepackage{lipsum}
\usepackage{graphicx}
\usepackage{subfigure}
\usepackage{sans}
\usepackage{changes}
\usepackage{hyperref}
\hypersetup{colorlinks=true,linkcolor=blue,urlcolor=blue,citecolor=blue}
\usepackage[toc,page]{appendix}
\usepackage[normalem]{ulem}
\usepackage{adjustbox}
\usepackage{latexsym}
\usepackage{amsmath}
\usepackage{amssymb}
\usepackage{amsfonts}
\usepackage{dcolumn}
\usepackage{bm}
\usepackage{tikz}
\usepackage{bigints}
\usepackage{array,tabularx,multirow,booktabs}
\usepackage[tracking=true]{microtype}
\SetTracking{}{500}
\SetTracking{encoding={*}, shape=sc}{40}
\UseRawInputEncoding 
\allowdisplaybreaks

\begin{document} \sloppy

\title{Shadow and greybody bounding of a regular scale-dependent black hole solution}


\author{Ali \"Ovg\"un
\orcidlink{0000-0002-9889-342X}
}
\email{ali.ovgun@emu.edu.tr}

\affiliation{Physics Department, Eastern Mediterranean
University, Famagusta, 99628 North Cyprus, via Mersin 10, Turkiye}

\author{Reggie C. Pantig
\orcidlink{0000-0002-3101-8591}
}
\email{rcpantig@mapua.edu.ph}
\affiliation{Physics Department, Map\'ua University, 658 Muralla St., Intramuros, Manila 1002, Philippines}

\author{\'Angel Rinc\'on
\orcidlink{0000-0001-8069-9162}
}
\email{angel.rincon@ua.es}
\affiliation{Departamento de Física Aplicada, Universidad de Alicante, Campus de San Vicente del Raspeig, E-03690 Alicante, Spain}

\date{\today}
\begin{abstract}
In this manuscript, we explore the shadow and the greybody bounding characteristics of a regular black hole within 4-dimensional space-time, employing the context of gravity that is scale-dependent. Our focus lies in determining limitations on the parameter denoted as $\tilde{\epsilon}$, which serves as a descriptor for the scale-dependent solution with respect to the classically observed shadow radius $R_\text{sh}$ as indicated in the documented data from the Event Horizon Telescope (EHT). Our result indicates that there is a unique value for $\tilde{\epsilon}$ occurring at the reported mean of $R_\text{sh}$ and the uncertainties could be the result of the fluctuating value of the scale-dependent parameter. We found that $\tilde{\epsilon} > 0$ is positive for Sgr. A*, but $\tilde{\epsilon}<0$ for M87*. Utilizing M87* as a reference model, we scrutinized the shadow radius and weak deflection angle within the specified constraints. Discrepancies were observed not only in the shadow but also in the deflection angle of photons, particularly when the photon's impact parameter closely approached the critical impact parameter.

\end{abstract}

\keywords{General relativity; Black holes; Shadow; Deflection angle; Greybody}

\pacs{95.30.Sf, 04.70.-s, 97.60.Lf, 04.50.+h}

\maketitle
\section{Introduction} \label{intro}
The existence of black holes, regions of spacetime where the gravitational pull is so strong that nothing, not even light, can escape, is a consequence of Einstein's theory of general relativity. There are evidence which support that BHs are more than just simple solutions of Einstein field equations. One of the most remarkable examples are the gravitational waves, confirmed a few years ago \cite{LIGOScientific:2016wkq,LIGOScientific:2016vlm}.
Albeit black holes have a long history, the first representative example was provided by Schwarzschild more than 100 years ago \cite{Schwarzschild:1916uq}. The discovery of black holes highlighted the limitations of Newtonian physics in describing gravity and underscored the profound implications of Einstein's general theory of relativity. Even in the absence of matter, Einstein's equations give rise to non-trivial solutions, such as black holes, whose properties deviate significantly from those of flat Minkowski space-time.
Black holes are fascinating by themselves since the physics required to understand them combines classical and quantum aspects. A significant breakthrough in black hole physics was Stephen Hawking's groundbreaking work \cite{Hawking:1974rv,Hawking:1975vcx}, which demonstrated that black holes emit radiation from their event horizons. This discovery made black holes a valuable laboratory for exploring and comprehending the nuances of gravitational theories.
The interior of black holes remains shrouded in uncertainty due to the presence of singularities, which are predicted to occur under certain conditions \cite{Hawking:1973uf}. Under general assumptions about the energy-matter content of spacetime, classical solutions to Einstein's equations exhibit both future singularities, as described by Penrose \cite{Penrose:1964wq}, and past singularities, as identified by Hawking \cite{Hawking:1965mf,Hawking:1966sx,Hawking:1966jv,Hawking:1967ju}. These singularities are typically concealed behind an event horizon \cite{Israel:1967za}.
However, there exist instances where static black hole solutions possess an event horizon yet the corresponding curvature invariants,  $R$, $R_{\mu\nu}R^{\mu\nu}$, $R_{\kappa\lambda\mu\nu}R^{\kappa\lambda\mu\nu}$  remain finite throughout the entire range of the radial coordinate $r$. These cases are commonly referred to as regular (or nonsingular) black holes (see \cite{Dymnikova:1992ux, Nicolini:2005vd, Hayward:2005gi, Xiang:2013sza, Ghosh:2014pba, Frolov:2016pav, Frolov:2017rjz} and references therein). These electrically neutral black holes are characterized by their mass and an additional parameter, and they asymptotically resemble the Schwarzschild solution. By introducing a non-zero electric (or magnetic) charge, we can identify notable examples of regular black holes. However, unlike the Reissner-Nordstrom solution, their behavior deviates in the weak field limit. 
Additional non-trivial examples will be discussed below.

The Bardeen solution~\cite{Bardeen:1968}, which can be derived by assuming either nonlinear electrodynamics for a magnetic source~\cite{Ayon-Beato:2000mjt} or an electric source~\cite{Rodrigues:2018bdc}. The Hayward solution, as presented in Ref.~\cite{Hayward:2005gi}. 
Similarly, the Hayward solution can be derived from a charged solution of this kind~\cite{Balart:2014jia}. Therefore, a common approach to resolving the singularity issue in black holes is to incorporate nonlinear electrodynamics into the action, as proposed in Refs.~\cite{Bronnikov:2017sgg,He:2017ujy}. While this approach has been successful in producing regular black hole solutions, it is important to note that the underlying physics of nonlinear electrodynamics remains largely unexplored.
In this respect, several well-known solutions have been investigated in detail, starting from the Nariai black hole solution \cite{1950SRToh..34..160N}, passing by the Ayon-Beato $\&$ Garcia regular solution \cite{Ayon-Beato:2000mjt}, including Balart $\&$ Vagenas \cite{Balart:2014cga} case, etc... Also, let us reinforce that the singularity problem could be reinterpreted as a sign that we must go beyond General Relativity to adequately describe the physics behind a black hole. In such a sense, a singularity could reveal the underlying necessity of extending GR due to including quantum features.

The shadow of a black hole is always worth the investigation since the distortion of spacetime continuum may leave imprints of the astrophysical environment that may affect it. The idea originates first in the seminal study of Synge \cite{Synge:1966okc} pertaining the photon escape. Then, Luminet \cite{Luminet:1979nyg} studied extensively the image of the black hole with thin accretion disk due to escaping photons. Just recently, the supermassive black hole images of M87* and Sgr. A* were announced in public \cite{EventHorizonTelescope:2019dse,EventHorizonTelescope:2022xnr}, showing how the electromagnetic spectrum in combination with impeccable data analysis techniques reveals that black holes are real objects in the Universe. The image shows the silhouette of the event horizon (dark zone), while the \textit{invicible classical} shadow boundary (which is $42\mu$as for M87* and $48.7\mu$as for Sgr. A*) was enveloped with glowing accretion disks \cite{Dokuchaev:2019jqq}. The paper is concerned only with the classical shadow as always done in the literature. In particular, we use the formalism developed in \cite{Perlick:2015vta,Perlick:2021aok}, which is enough to study the shadow behavior of non-rotating black holes. The method has become successful in studying the effects of various astrophysical environments on the black hole geometry, as well as those spacetime metrics involving scale-dependencies and quantum corrections. See Refs. \cite{Afrin:2021imp,Atamurotov:2015nra,Atamurotov:2015xfa,Cunha:2018acu,Cunha:2016wzk,Herdeiro:2021lwl,Kuang:2022xjp,Kuang:2022ojj,Meng:2022kjs,Lima:2021las,Pantig:2022toh,Lobos:2022jsz,Rayimbaev:2022hca,Pantig:2022ely,Pantig:2022whj,Pantig:2022sjb,Contreras:2019nih,Contreras:2019nih,Khodadi:2022ulo,Badia:2022phg,Pantig:2022gih,Pantig:2022qak,Atamurotov:2022wsr,Contreras:2019cmf,Lambiase:2022ywp,Kumar:2020hgm,Kumar:2018ple,Saghafi:2022pme,Guo:2022nto,Cimdiker:2021cpz,Xu:2021xgw,Konoplya:2022hbl,Atamurotov:2013sca,Konoplya:2019sns,Sokoliuk:2022owk,Ovgun:2023ego,Pantig:2023yer,Ovgun:2018tua,Uniyal:2022vdu,Vagnozzi:2019apd,Khodadi:2021gbc,Roy:2020dyy} to cite a few.

In 1916, Albert Einstein made a remarkable prediction about gravitational lensing in his revolutionary theory of general relativity. This prediction suggested that massive objects, such as stars and galaxies, would bend the fabric of spacetime, causing light rays to deviate from their expected paths. This phenomenon is known as gravitational lensing.

In 1919, British astronomer Sir Arthur Eddington provided the first experimental evidence of gravitational lensing during a total solar eclipse. Eddington measured the deflection of starlight passing near the Sun and found that it matched Einstein's prediction with remarkable accuracy. This observation marked a watershed moment in the history of physics, confirming the validity of general relativity and establishing Einstein as one of the greatest scientists of all time. Since Eddington's groundbreaking observation, gravitational lensing has become an invaluable tool in astrophysics, enabling astronomers to probe the depths of the cosmos and uncover the mysteries of the universe. Gravitational lensing has been used to study a wide range of phenomena, including: Dark matter where gravitational lensing has provided strong evidence for the existence of dark matter, an invisible substance that makes up about $85\%$ of the matter in the universe. Dark matter's gravitational pull can bend light rays, revealing the presence of dark matter even though it cannot be directly observed \cite{Virbhadra:1999nm,Virbhadra:2002ju,Adler:2022qtb,Bozza:2001xd,Bozza:2002zj,Perlick:2003vg,Virbhadra:2022ybp,Virbhadra:2022iiy}. Gravitational lensing has been used to detect and study exoplanets, planets orbiting stars other than our Sun. By observing the subtle bending of light from a distant star as it passes near an exoplanet, astronomers can infer the planet's mass and orbit. Gravitational lensing has been used to study the expansion of the universe, providing valuable insights into the universe's past, present, and future. By measuring the distances to distant objects using gravitational lensing, astronomers have been able to confirm that the expansion of the universe is accelerating. Over the years, numerous studies have explored the relationship between weak deflection angle and the Gauss-Bonnet theorem (GBT). The GBT is a fundamental theorem in differential geometry that relates the curvature of a surface to its topology. In the context of gravitational lensing, the GBT can be used to calculate the deflection of light rays in static, asymptotically flat spacetimes.

In 2008, Gibbons and Werner demonstrated a method for calculating the deflection angle using the GBT for asymptotically flat static spacetimes \cite{Gibbons:2008rj}. This method opened up new possibilities for studying gravitational lensing in various astrophysical scenarios. Subsequently, Werner extended this method to stationary black holes \cite{Werner:2012rc}, and Ishihara et al. showed that deflection angles could be calculated for finite distances \cite{Ishihara:2016vdc}. These developments further enhanced the applicability of the GBT in gravitational lensing studies. The growing interest in weak gravitational lensing via the GBT methods stems from its potential to provide a deeper understanding of gravitational lensing phenomena in black holes, wormholes, and other complex spacetimes. By applying the GBT to these systems, astronomers can gain valuable insights into the behavior of light and matter in the vicinity of these highly curved regions of spacetime \cite{Takizawa:2020egm,Ono:2019hkw,Ono:2017pie,Asada:2017vxl,Ovgun:2018fnk,Ovgun:2019wej,Ovgun:2018oxk,Ovgun:2020gjz,Ovgun:2018prw,Ovgun:2018ran,Javed:2019ynm,Li:2020dln,Li:2020wvn,Belhaj:2022vte,Pantig:2022toh,Javed:2023iih,Javed:2020lsg,Javed:2022psa,Javed:2022fsn,Javed:2022gtz,Ovgun:2020yuv,Javed:2019jag,Javed:2019rrg,Kumaran:2019qqp,Jusufi:2017mav,Javed:2019qyg}.

The greybody factor of a black hole is a crucial parameter that quantifies the probability of a particle or wave being absorbed or scattered by the black hole. This factor depends on the energy and angular momentum of the incident particle or wave, as well as the properties of the black hole, such as its mass, spin, and charge. The greybody factor plays a significant role in understanding the behavior of matter and radiation in the vicinity of black holes and is an essential component in determining their observed emission and absorption spectra.

The presence of event horizons, the boundaries around black holes beyond which nothing, not even light, can escape, has a profound impact on the greybody factor. At low energies, the event horizon suppresses the greybody factor, leading to a phenomenon known as the Hawking effect. This effect, predicted by Stephen Hawking in 1975, suggests that black holes are not entirely black but emit radiation, now known as Hawking radiation \cite{Hawking:1975vcx}.  Greybody factors have been extensively studied in the context of black hole thermodynamics and quantum gravity, providing valuable insights into the nature of these enigmatic objects. By studying the behavior of greybody factors, physicists can gain a deeper understanding of the processes that occur near black holes and the relationship between gravity and quantum mechanics \cite{Parikh:1999mf,Singleton:2011vh,Akhmedova:2008dz}. Several methods have been developed to calculate greybody factors, each with its own strengths and limitations \cite{Maldacena:1996ix,Cvetic:1997uw,Harmark:2007jy,Gubser:1996zp,Zhang:2020qam,Fernando:2004ay,Kanti:2014dxa,Okyay:2021nnh,Panotopoulos:2018pvu,Panotopoulos:2016wuu,Rincon:2018ktz,Panotopoulos:2017yoe,Ahmed:2016lou,Al-Badawi:2022aby,Al-Badawi:2021wdm,Javed:2021ymu,Javed:2022kzf,Javed:2022rrs,Lee:1998pd,Pappas:2016ovo,Crispino:2013pya,Chen:2010ru,Devi:2020uac,Gogoi:2022wyv,Kanti:2009sn}. Some common methods include: Maldacena's Conformal Field Theory (CFT) Approach: This method utilizes the AdS/CFT correspondence, a powerful tool in theoretical physics, to relate the greybody factors of black holes in higher-dimensional spacetimes to correlators in lower-dimensional conformal field theories \cite{Maldacena:1996ix,Cvetic:1997uw,Harmark:2007jy,Gubser:1996zp,Zhang:2020qam}. Another method 's Quasinormal Mode Approach: This approach relies on the concept of quasinormal modes, resonant frequencies that characterize the response of a black hole to external perturbations \cite{Zhang:2020qam,Fernando:2004ay,Kanti:2014dxa}. By studying the quasinormal modes, researchers can determine the greybody factors for specific types of particles or waves. Another method is WKB Approximation: This semi-classical method approximates the solutions to wave equations in curved spacetimes, allowing for the calculation of greybody factors in certain cases \cite{Ovgun:2018gwt,Konoplya:2003ii,Berti:2009kk,Konoplya:2011qq,Zhidenko:2005mv,Daghigh:2008jz,Daghigh:2011ty} . In 1998, Matt Visser introduced an elegant analytical method to derive rigorous bounds on greybody factors \cite{Visser:1998ke,Boonserm:2008zg}. This method has been further refined by Boonserm and others, providing valuable constraints on the possible values of greybody factors for various black hole configurations. By analyzing the greybody factors emitted by black holes, astronomers can infer the mass, spin, and charge of these objects.  Greybody factors provide insights into how particles and waves interact with the intense gravitational fields surrounding black holes.  The study of greybody factors contributes to our understanding of quantum gravity, the theoretical framework that seeks to reconcile quantum mechanics with general relativity \cite{Boonserm:2017qcq,Boonserm:2019mon,Liu:2022ygf,Yang:2022ifo,Gray:2015xig,Boonserm:2014fja,Boonserm:2014rma,Boonserm:2013dua,Ngampitipan:2012dq,Boonserm:2009zba,Boonserm:2009mi,Lambiase:2023zeo,Gogoi:2023fow}.

In this paper, we delve into the intriguing realm of regular black holes within the framework of scale-dependent gravity. This alternative approach to general relativity seeks to incorporate corrections to the classical background by introducing scale-dependent couplings, drawing inspiration from quantum gravity theories. 
What is more, by incorporating scale-dependent couplings, we can potentially resolve the singularity issue and obtain a more comprehensive description of black hole physics.
Our motivation stems from the recent groundbreaking observations of black hole images from M87* and Sgr. A*, which has opened up new avenues for understanding these enigmatic celestial objects.
Specifically, we investigate the black hole shadow, a crucial feature that reveals the gravitational influence of a black hole on its surroundings. By analyzing the shadow cast by a regular black hole in scale-dependent gravity, we aim to establish constraints on its parameters and gain insights into its behavior.
Furthermore, we explore the greybody bounding of this regular black hole solution. Greybody factors quantify the probability of a particle or wave being absorbed or scattered by the black hole. By studying the greybody factors, we can gain valuable information about the interactions between matter and radiation in the vicinity of black holes.


The present manuscript is organized as follows: after this concise introduction, we will briefly mention the scale-dependent formalism, its equations and the corresponding black hole solution (to be studied in Sect. \ref{sec2}).
Subsequently, we find constraints for $\tilde{\epsilon}$ using the EHT data to analyze properly the possibility of its deviational effects relative to the Schwarzschild case. 
In particular, we will investigate the impact of this scale-dependent parameter to the photonsphere radius, and shadow radius using Sgr. A* and M87* as models (in Sect. \ref{sec3}). 
In Sect. \ref{sec4} we will study the weak deflection angle of the black hole.  
After that, in Sect. \ref{sec5}, we will calculate the greybody bounding of the new solutions. 
Our results are presented in figures for better comprehension. We used natural units as $G = c = 1$ and metric signature $(-,+,+,+)$.

\section{Scale-dependent gravity}\label{sec2}
%
In the present section, we will summarize just the main ingredients needed to understand the scale-dependent scenario, an approach substantially based on asymptotically safe gravity, i.e., a theory that combines classical and quantum aspects in a self-consistent way.  
The interest in gravitational theories where classical and quantum physics work well is profound \cite{Padmanabhan:2001ev,Padmanabhan:1998yy}. 
Firstly, a perfect example of that emerges via the study of black holes. They represent the perfect arena in which gravity and quantum features are mixed, which suggest that classical and quantum aspect are related each other, and, second, due we are still looking for a complete theory to describe the universe irrespectively of the scale. 

Thus, focusing on quantum-inspired theories of gravity, the implementation of that on black hole physics plays a prominent role. Why? Because a wide variety of cases in which the inclusion of quantum features on a classical background has been consistently implemented.
In particular, there are at least three different ways to modify classical solutions. We can split them as follows \cite{Reuter:2003ca,Rincon:2022hpy}:
  i) at the level of solution,
 ii) at the level of the equations of motion, and
iii) at the level of the action.
One of the most remarkable examples of modification of general relativity to account for quantum features is the seminal paper of Bonanno and Reuter \cite{Bonanno:2000ep}. That work contains a detailed study of how the renormalization group effects could disturb the classical Schwarzschild black hole solution.
Thus, “renormalization group improving” classical
metrics, or better known as improved black hole solutions, take as critical ingredient the existence of a running/scale-dependent Newton coupling ``constant" which is, in fact, obtained from the exact evolution equation for the effective average action. 
Based on that successful work, an alternative research line in which general relativity was modified emerged (see  \cite{Koch:2014cqa,Cai:2010zh,Gonzalez:2015upa,Platania:2023srt,Ishibashi:2021kmf} to name a few). 

To maintain the discussion compact, we will mention the essentials here;
the fundamental object used to parameterize the theory (irrespective of the physics we will consider) is the average effective action $\Gamma[g_{\mu\nu}, k, \cdots]$. Such an object replaces the classical action because the classical couplings are now functions that depend on the arbitrary scale $k$. Such scale is, in practice, a function of coordinates $x^{\mu}$. Thus, taking advantage of concrete symmetry, we expect $k$ evolves only as a  function of a single coordinate. For the spherically symmetric case, it is expected that $k(x^{\mu}) = k(r)$. Finally, with that in mind, the set of functions considered initially as $\{ G_k, \Lambda_k \}$, can be treated as $\{ G(r), \Lambda(r) \}$, simplifying the problem significantly. 

The way to proceed here is i) to compute the effective Einstein's field equations by taking the corresponding variations to the metric from the effective action and ii) to compute the consistency condition by taking the variation to scale $k$. The last part is not easy, and, in general, the resulting equation does not produce an analytical solution. Thus, an NEC-like condition can circumvent that problem, in agreement with many original papers. Finally, in the next subsection we elaborate a little bit regarding how to obtain the equation which conduce to this regular black hole in scale-dependent gravity.

\subsection{A regular scale-dependent black hole (RSDBH)} 
In order to properly describe the problem, we will start by considering an average effective action, $\Gamma[g_{\mu\nu},k]$, with three ingredients only: 
i) the Ricci scalar, $R$, 
ii) the cosmological parameter, $\Lambda_k$, and 
iii) the matter action, $S_k$. 
Also, the classical Einstein coupling, $\kappa_0$, is replaced by its scale-dependent counterpart $\kappa_k=8\pi G_k$, being $G_k$ the scale-dependent Newton's coupling. 
Thus, following the original paper \cite{Contreras:2017eza}, the scale-dependent action is then given by
\begin{equation}
\Gamma[g_{\mu\nu},k]=\int d^4 x \sqrt{-g}\left[\frac{1}{2\kappa_k}(R-2\Lambda_k)\right]+S_k,
\end{equation}
where $c=1$, and $\Lambda_k$ is the cosmological couplings respectively. 
Notice that the scale-dependent effect is encoded into the functions via the index $k$.
To obtain the equations of motion, we can proceed in a conventional way, i.e., we first obtain a set of Einstein-like equations, and, subsequently, we close the system utilizing variations of the effective action with respect to the renormalization scale $k$, or, by means of another supplementary condition, for instance, an energy-like constraint.

Firstly, taking the corresponding variation with respect to the inverse metric field $g^{\mu \nu}$, we obtain the modified form of Einstein's field equations, i.e., 
\begin{equation}
G_{\mu\nu}+\Lambda_k g_{\mu\nu}=\kappa_k (T^{\textrm{eff}})_{\mu\nu},
\label{einstein}
\end{equation}
being $(T^{\textrm{eff}})_{\mu\nu}$ the corresponding effective energy-momentum tensor, defined according to
\begin{equation}
(T^{\textrm{eff}})_{\mu\nu}:=(T_{\mu\nu})_k-\frac{1}{\kappa_k}\Delta t_{\mu\nu},
\end{equation}
where we have introduced an auxiliary tensor $\Delta t_{\mu\nu}$, which accounts for the running of the gravitational Newton's coupling
\begin{equation}
\Delta t_{\mu\nu}=G_k
\Bigl(
g_{\mu\nu}\,\square-\nabla_\mu \nabla_\nu
\Bigl)
G^{-1}_{k}.
\end{equation}
Secondly, taking into account the scale setting $k \rightarrow k(x)$,
the variational approach \cite{Koch:2014joa}
could be useful, because it guarantees a minimal dependence on the arbitrary renormalization parameter $k$
\begin{align}\label{scale}
\frac{\mathrm{d}}{\mathrm{d}k}\Gamma[g_{\mu \nu}, k] =0.
\end{align}
Roughly speaking, combining Eq. \eqref{einstein} with the equation obtained from \eqref{scale},
we could close the system and, therefore, obtain a new solution. Also, notice that the scale setting \eqref{scale} represents a way to determine the scalar function $k(x)$.
Replacing such a solution back into $\Gamma_k$, would (up to a boundary term) give the effective action, obtaining an agreement with the corresponding symmetries.
Albeit the latter is formally true, it is also true that there is a
lack reliable of knowledge of $\Gamma_k$ (coming from the concrete form of beta functions of quantum gravity), which basically frustrates such attempts.
In view of that, an alternative strategy could provide us with a more suitable form the face the problem. If we accept that classical couplings $\{\alpha_0, \beta_0, \gamma_0, \cdots_0 \}$ now evolve with respect to the arbitrary scale $k$, we can replace such a set by its corresponding scale-dependent or running counterpart $\{\alpha_k, \beta_k, \gamma_k, {(\cdots)}_k \}$.

This problem can be circumvented by promoting both $G$ and $\Lambda$ to field variables and by imposing one additional constraint. Usually, the best choice to close the system is the saturated version of the Null Energy Condition (NEC) widely used over the years.
%
%
In particular, by demanding the saturated version of the NEC on $(T^{eff})_{\mu\nu}$, we can close and solve consistently the problem. Thus, considering the effective energy-momentum tensor, $(T^{eff})_{\mu\nu}$, the above-mentioned condition can be written as
\begin{eqnarray}
(T^{eff})_{\mu\nu} \ell^{\mu}\ell^{\nu}= \left[T_{\mu\nu} - \frac{1}{\kappa(r)}\Delta t_{\mu\nu}\right]\ell^{\mu}\ell^{\nu} \ge 0,
\end{eqnarray}
where $\ell^{\mu}$ is a radial null vector. 
For simplicity, we can set $A(r) =  f(r)$ and $B(r) = f(r)^{-1}$, albeit the discussion is still valid in more complicated cases. So, let us considering the case $\ell^{\mu}=\{ f^{-1/2}, f^{1/2} , 0, 0 \}$. The contraction $T_{\mu\nu}\ell^{\mu}\ell^{\nu}=0$ implies that $\Delta t_{\mu\nu}\ell^{\mu}\ell^{\nu}\ge 0$.
In addition, it is obvious that that $G_{\mu\nu}\ell^{\mu}\ell^{\nu}=0$ and, for consistency with Eq. (\ref{einstein}), we
demand
\begin{eqnarray}\label{necnm}
\Delta t_{\mu\nu}\ell^{\mu}\ell^{\nu}=0.
\end{eqnarray}
and solving the previous equation we obtain the conventional form of the gravitational coupling, $G(r)$, as was
previously mentioned in Ref. \cite{Rincon:2017ayr}. 
To be more precise, the differential equation for the gravitational coupling is then given by 
\begin{align}\label{EDO_G}
2\left[\frac{\mathrm{d} G(r)}{\mathrm{d} r}\right]^2 = G(r)\frac{\mathrm{d}^2 G(r) }{\mathrm{d} r^2 },
\end{align}
and the corresponding solution for $G(r)$ is then
\begin{eqnarray}\label{gr}
G(r)=\frac{G_{0}}{1+\epsilon r},
\end{eqnarray}
At this point, we should mention that, as we have a second-order differential equation, we have two integration constants to fix. The first is in the so-called classical Newton coupling, $G_0$, which is fixed in such a way that in some limit we can always recover the classical case. The second constant, $\epsilon$, is the so-called scale-dependent parameter, which has dimensions equal to the inverse of the length, and plays an important role because it allows us to account for the effect of the quantum feature on the classical background. Thus, when $\epsilon$ goes to zero, we recover the standard/classical solution (coming from GR), whereas when we turn $\epsilon$ on, we obtain quantum-like solutions. At this point, it is clear that $\epsilon$ can take both positive and negative values, although theoretically positive values are preferable, but both are acceptable.
Nevertheless, 
there are several examples where the quantum effect can take both positive and negative values. Some examples are
\cite{Lambiase:2022xde,Scardigli:2022jtt,Lambiase:2023hng}. Note that these last examples are NOT in the context of scale-dependent gravity, but we have included them to illustrate that the quantum correction can also be positive or negative. The same happens in the concrete case of scale-dependent gravity, because the theory accepts both situations ($\epsilon > 0$ or $\epsilon < 0$), or more precisely, the theory does not forbid any value of $\epsilon$.
Additional details regarding the application of scale-dependence in 
  i)  black hole physics (and their properties) 
 ii)  wormholes 
iii) relativistic stars and
 iv) cosmological models 
can be consulted in 
\cite{Rincon:2017goj,Rincon:2018sgd,Rincon:2018lyd,Rincon:2019zxk,Fathi:2019jid,Contreras:2018gpl,Rincon:2020cpz,Rincon:2019cix,Sendra:2018vux,Panotopoulos:2021tkk,Contreras:2018swc,Canales:2018tbn,Alvarez:2022wef,Panotopoulos:2021heb,Alvarez:2022mlf,Alvarez:2020xmk,Bargueno:2021nuc,Panotopoulos:2021obe,Panotopoulos:2020zqa} and references therein.
However, let us highlight that scale-dependent gravity produces alternative solutions from other similar quantum-inspired approaches, like asymptotically safe gravity. In particular, should be mentioned that, as was pointed out in \cite{Rincon:2019cix}, scale-dependent gravity generally gives opposite results in comparison to asymptotically safe gravity (see Fig.[4] in Ref.\cite{Rincon:2019cix} and discussion about that).
However, there exist a case in which scale-dependent gravity is consistent with some results coming from asymptotically safe gravity, which is with the so-called "infrared instability" (see \cite{Biemans:2016rvp}). 
Scale-dependent gravity also has some implications at the cosmological level, for instance, the fact that the cosmological constant problem and the $H_0$ problem can be alleviated, which is, of course, a tremendous advantage and also makes more attractive the study of alternative theories of gravity (see for instance \cite{Alvarez:2020xmk,Bargueno:2021nuc,Panotopoulos:2021heb} for more details).
Finally, and just to reinforce the idea of scale-dependent gravity, we should mention that for large $r$ we have a scale-dependent weak gravity, while at small distances we have a modified version of Newtonian gravity.

Considering Eq. (\ref{einstein}) and assuming a spherically symmetric and static geometry, the following metric is obtained:\cite{Contreras:2017eza}
\begin{equation}
ds^{2}=-A(r)dt^{2}+B(r)dr^{2}+C(r)(d\theta ^{2}+\sin ^{2}\theta d\phi ^{2}),  \label{m1}
\end{equation}
where
\begin{equation}
A(r)=B(r)^{-1}=1-\frac{2MG_0}{r}\left(1+\frac{M^2 G^{2}_{0}\epsilon}{6r}\right)^{-3}, \hspace{1cm} C(r)=r^2,
\label{metricelement}
\end{equation}
with $M$  representing the mass, $G_0$ representing Newton's gravitational constant, and $\epsilon>0$ being a positive running parameter with dimensions of inverse length associated with the scale-dependent gravitational coupling $G_k$. The metric asymptotically approaches the Schwarzschild metric as r approaches infinity: $A(r)\rightarrow 1-2G_0 M/r$ as $r\rightarrow\infty$. Moreover, the metric is regular everywhere (non-singular) and exhibits a de Sitter behavior for $r\rightarrow 0$: since $A(r)\rightarrow 1-432r^2/G^{5}_{0}M^5\epsilon^3$. In the limit $r\rightarrow 0$, an effective cosmological constant $\Lambda_{\textrm{eff}}=1296/G^{5}_{0}M^{5}_{0}\epsilon^3$ emerges. This regular solution was obtained without resorting to nonlinear electrodynamics and corresponds to a semi-classical extension of the Schwarzschild black hole. It can be interpreted without the cosmological term by attributing the matter content to an anisotropic vacuum that modifies the usual Schwarzschild geometry. This vacuum energy can be considered to be of quantum nature and is non-negligible for non-zero values of $\epsilon$, where the scale dependence effect becomes prominent. In the limit $\epsilon\rightarrow 0$, the Schwarzschild solution is recovered. There are no definitive constraints on the running parameter $\epsilon$, but since it arises from quantum corrections, it is expected to assume small values. As a final remark about approximations, we find that when $r\rightarrow \infty$, a third term correction as $\epsilon M^3 / r^2$ arises. In this far-field approximation, we see that this is identical to the correction found in \cite{Scardigli:2014qka} (see Eq. 29) if we define $\epsilon' = M \epsilon$.

The metric (\ref{m1}) can be conveniently rewritten by defining the quantities   can be conveniently written by defining the quantities $T=t/M$, $x=r/M$ and $\tilde{\epsilon}=\epsilon M$, and adopting $G_0=1$: \cite{Sendra:2018vux}
\begin{equation}
ds^{2}=-A(x)dT^{2}+B(x)dx^{2}+C(x)(d\theta ^{2}+\sin ^{2}\theta d\phi ^{2}),  \label{m2}
\end{equation}
where
\begin{equation}
A(x)=B(x)^{-1}=1-\frac{2}{x}\left(1+\frac{\tilde{\epsilon}}{6x}\right)^{-3}, \hspace{1cm} C(r)=x^2.
\label{metricelement2}
\end{equation}
\begin{figure*}
    \centering
    \includegraphics[width=0.48\textwidth]{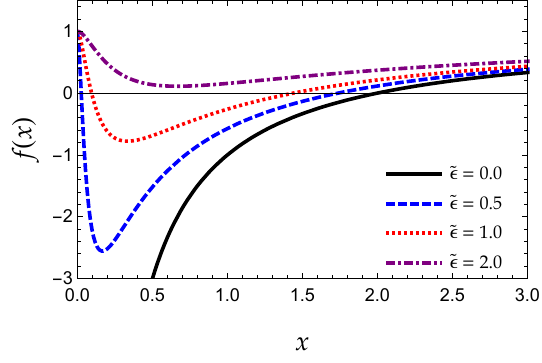}
    \includegraphics[width=0.48\textwidth]{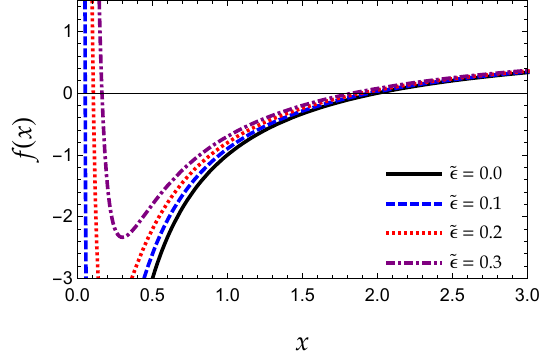}
    \caption{Lapse function $f(x)$ against the dimensionless radial variable $x$, assuming $G_0 =1$ and $M=1$, for different values of the scale-dependent parameter $\epsilon$.
    {\bf{Left panel}}: Exact lapse function 
    {\bf{Right panel}}: approximated lapse function
    }
    \label{lapse}
\end{figure*}

The event horizon radius, the boundary beyond which nothing, not even light, can escape the black hole's gravitational pull, is determined by setting the metric function $A(x)=0$ equal to zero. This yields the following equation:
\begin{equation}
216~x^{3} - \left(432 - 108 \tilde{\epsilon}\right)x^{2} + 18~\tilde{\epsilon}^2 x + \tilde{\epsilon}^3 = 0,
\label{xh}
\end{equation}
and corresponds to a third-degree polynomial equation, where the solutions are:
\begin{align} \label{xh2}
x_1 &=\frac{1}{6} \Bigg\{4 - \tilde{\epsilon} + \frac{2^{8/3} (2 - \tilde{\epsilon})}{\left[32 - \tilde{\epsilon} \left(24 - 3 \tilde{\epsilon} - \sqrt{9  \tilde{\epsilon}^2-16 \tilde{\epsilon}}\right)\right]^{1/3}}+2^{1/3} \left[32 - \tilde{\epsilon} \left(24 - 3 \tilde{\epsilon} - \sqrt{9 \tilde{\epsilon}^2-16 \tilde{\epsilon}}\right)\right]^{1/3}\Bigg\} ,
\\
x_2 &=
\frac{1}{12} \left(2 (-1)^{2/3} \sqrt[3]{2 \tilde{\epsilon } \left(3 \tilde{\epsilon }+\sqrt{\tilde{\epsilon } \left(9 \tilde{\epsilon }-16\right)}-24\right)+64}-2 \tilde{\epsilon }+\frac{8 \sqrt[3]{-1} 2^{2/3} \left(\tilde{\epsilon }-2\right)}{\sqrt[3]{3 \left(\tilde{\epsilon }-8\right) \tilde{\epsilon }+\sqrt{\tilde{\epsilon }^3 \left(9 \tilde{\epsilon }-16\right)}+32}}+8\right) ,
\\
x_3 &= 
\frac{1}{12} \left(-2 \sqrt[3]{-2} \sqrt[3]{\tilde{\epsilon } \left(3 \tilde{\epsilon }+\sqrt{\tilde{\epsilon } \left(9 \tilde{\epsilon }-16\right)}-24\right)+32}-2 \tilde{\epsilon }-\frac{8 (-2)^{2/3} \left(\tilde{\epsilon }-2\right)}{\sqrt[3]{3 \left(\tilde{\epsilon }-8\right) \tilde{\epsilon }+\sqrt{\tilde{\epsilon }^3 \left(9 \tilde{\epsilon }-16\right)}+32}}+8\right) ,
\end{align}
being the black hole horizon $x_h = \text{max}\{ 
 x_1, x_2. x_3\}$.
At this point, it should be mentioned that the event horizon is smaller than its classical counterpart. We can see such an effect clearly by recognizing the Schwarzschild horizon ($r_0 \equiv 2 G_0 M$), we can consider the first and second-order corrections, i.e., 
\begin{align}
    r_1 &\approx r_0 
    \Bigg[ 
    1 - \frac{1}{8} (\epsilon r_0) - \frac{1}{192} (\epsilon r_0)^2 
    \Bigg] + \mathcal{O}(\epsilon^3),
    \\
    r_2 &\approx r_0 
     \Bigg[
     -\frac{1}{48\sqrt{6}}(\epsilon r_0)^{3/2} + \frac{1}{384}(\epsilon r_0)^2
     \Bigg] + \mathcal{O}(\epsilon^3),
     \\
    r_3 &\approx r_0 
     \Bigg[
     \frac{1}{48\sqrt{6}}(\epsilon r_0)^{3/2} + \frac{1}{384}(\epsilon r_0)^2
     \Bigg] + \mathcal{O}(\epsilon^3),
\end{align}
where, again $r_h = \text{max}\{ 
 r_1, r_2. r_3\}$.
\begin{figure*}
    \centering
    \includegraphics[width=0.48\textwidth]{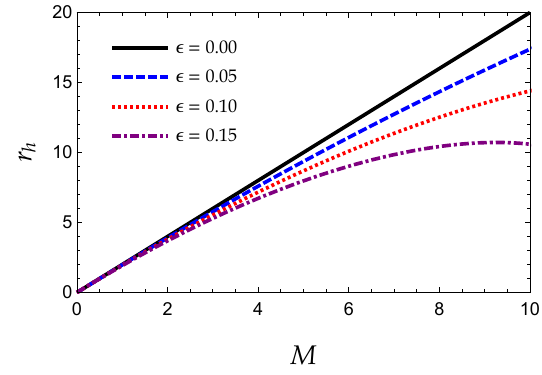}
    \includegraphics[width=0.48\textwidth]{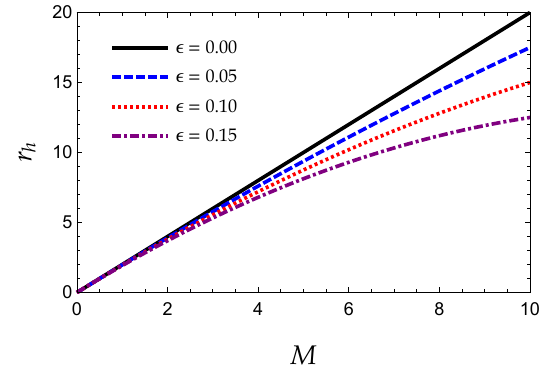}
    \caption{Black hole horizon $r_h$ against the classical mass $M$, assuming $G_0 =1$, for different values of the scale-dependent parameter $\epsilon$.
    {\bf{Left panel}}: Exact black hole horizon
    {\bf{Right panel}}: approximated black hole horizon
    }
    \label{horizon}
\end{figure*}

\section{Constraining $\tilde{\epsilon}$ through the shadow from EHT data} \label{sec3}
In this section, we will determine the bounds on $\tilde{\epsilon}$ through the phenomenon of black hole shadow and using the EHT data for Sgr. A* and M87*. Moreover, we will focus the perceived shadow radius by an observer located $(x,\theta,\phi) = (x_\text{obs}, \theta_\text{obs}, \phi_\text{obs})$, and we aim to examine how the shadow radius $R_\text{sh}$ varies with $x_\text{obs}$.

Without loss of generality, since the spacetime metric is spherically symmetric, the Lagrangian for light rays geodesics along the equatorial plane ($\theta = \pi/2$) is expressed as \cite{Perlick:2015vta,Perlick:2021aok}
\begin{equation}
{\mathcal{L}}(x , \dot{x} ) =
 \frac{1}{2} \left(- A(x) \dot{T}{}^2
+ B(x)\dot{x}{}^2 + C(x) \dot{\phi}{}^2 \right),
\end{equation}
where the overhead dot notation pertains to the derivative of a particular coordinate with respect to the affine parameter $\lambda$. With the variational principle, such a Lagrangian gives rise to two constants of motion (energy $E$ and angular momentum $L$ both in per unit mass)
\begin{equation} \label{econs}
E = A(x)   \dot{T}  , \qquad
L  =  C(x)  \dot{\phi}.
\end{equation}
Null geodesics requires that $ds^2=0$:
\begin{equation}
- A(x) \dot{T}{}^2  + 
B(x)\dot{x}{}^2  +  C(x) \dot{\phi}{}^2
 =  0,
\end{equation}
which now allows us to derive the orbit equation using Eq. \eqref{econs}
\begin{equation} \label{eq:orbit}
\left( \frac{dx}{d\phi} \right) ^2  = \frac{C(x)}{B(x)}\left(\frac{h(x)^2}{b^2} - 1 \right),
\end{equation}
where $h(x)^2 = C(x)/A(x)$, and $b$ is the impact parameter defined by
\begin{equation}
    b = \frac{L}{E} = \frac{C(x)}{A(x)} \frac{d\phi}{dT}
\end{equation}

The null circular orbit must satisfy the conditions $dx/d \phi =0$ for circularity, and $d^2x/d \phi ^2 =0$ for stability to find the photonsphere radius $x_\text{ps}$. Alternatively, the photonsphere radius can be found by satisfying the condition $h(x)' = 0$ where prime denotes derivative with respect to $x$. Our result gives
\begin{equation}
	1296{x}^{4}+ \left( 864\tilde{\epsilon}-3888 \right) {x}^{3}+216{x}^{2}{\tilde{\epsilon}}^{2}+24x{\tilde{\epsilon}}^{3}+{\tilde{\epsilon}}^{4}=0,
\end{equation}
where the roots determines the light ring radius $x_\text{ps}$. In the presence of small perturbations of photons undergoing circular motion at $x_\text{ps}$, the light rays can either spiral to the black hole or escape to infinity. For a static observer at $x_\text{obs}$ relative to the black hole's center, the escaping photons that will reach it are those that traveled the closest approach or the \textit{critical} impact parameter. For our metric, we define it as \cite{Pantig:2022sjb}
\begin{equation}
    b_\text{crit}^2 = \frac{4x_\text{ps}^2}{x_\text{ps} A'(x)|_{x=x_\text{ps}} +2A(x_\text{ps})}.
\end{equation}
Using Eq. \eqref{metricelement2}, we find as
\begin{equation}
    b_\text{crit}^2 = \frac{2 \left(6 x_\text{ps} +\tilde{\epsilon} \right)^{4} x_\text{ps}^{2}}{1296 x_\text{ps}^{4}+\left(864 \tilde{\epsilon} -1296\right) x_\text{ps}^{3}+\left(216 \tilde{\epsilon}^{2}-864 \tilde{\epsilon} \right) x_\text{ps}^{2}+24 x_\text{ps} \tilde{\epsilon}^{3}+\tilde{\epsilon}^{4}}.
\end{equation}

As $x_\text{obs}$ is considerably far from the black hole, a simple application of geometry will lead to the shadow's angular radius \cite{Perlick:2015vta}
\begin{equation} \label{eangshad}
\tan^2 \alpha_\text{sh}  =  
\frac{b_\text{crit}^2 A(x_\text{obs})}{C(x_\text{obs})}\frac{d \phi}{dx}
 \Bigg|_{x = x_\text{obs}}.
\end{equation}
Then, using the orbit equation in Eq. \eqref{eq:orbit}, and Eq. \eqref{eangshad}, the above can be written as
\begin{equation} \label{eqshad}
    R_\text{sh} = b_\text{crit}\left[1 - \frac{2}{x_\text{obs}}\left(1 + \frac{\tilde{\epsilon}}{6x_\text{obs}} \right)^{-3}    \right]^{1/2},
\end{equation}
which is the \textit{classical} shadow radius that is usually studied in the literature.

\begin{figure*}
    \centering
    \includegraphics[width=0.48\textwidth]{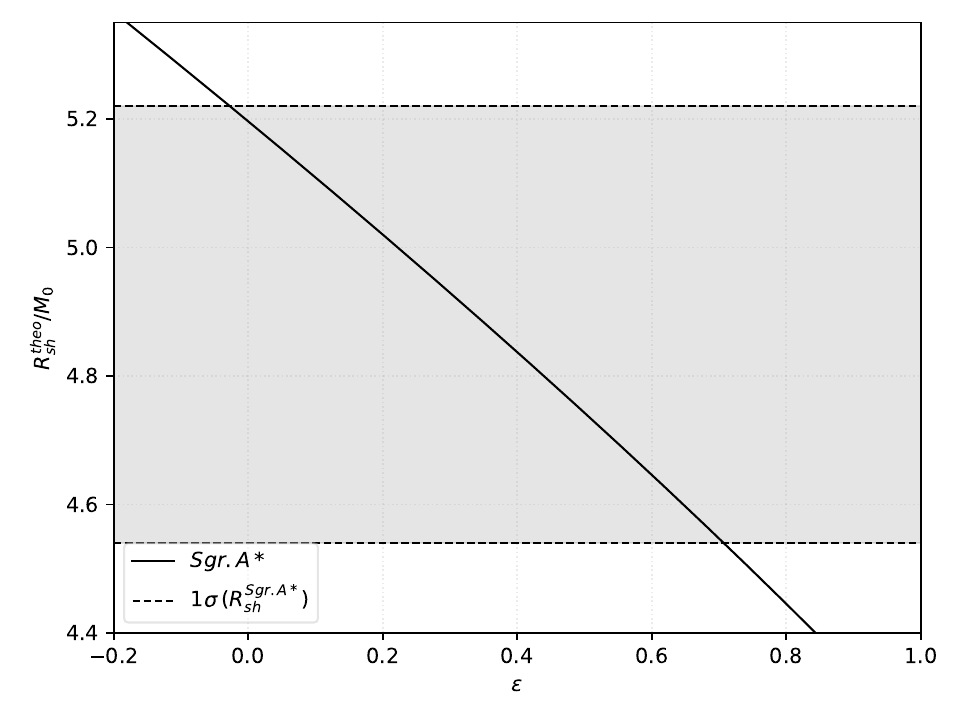}
    \includegraphics[width=0.48\textwidth]{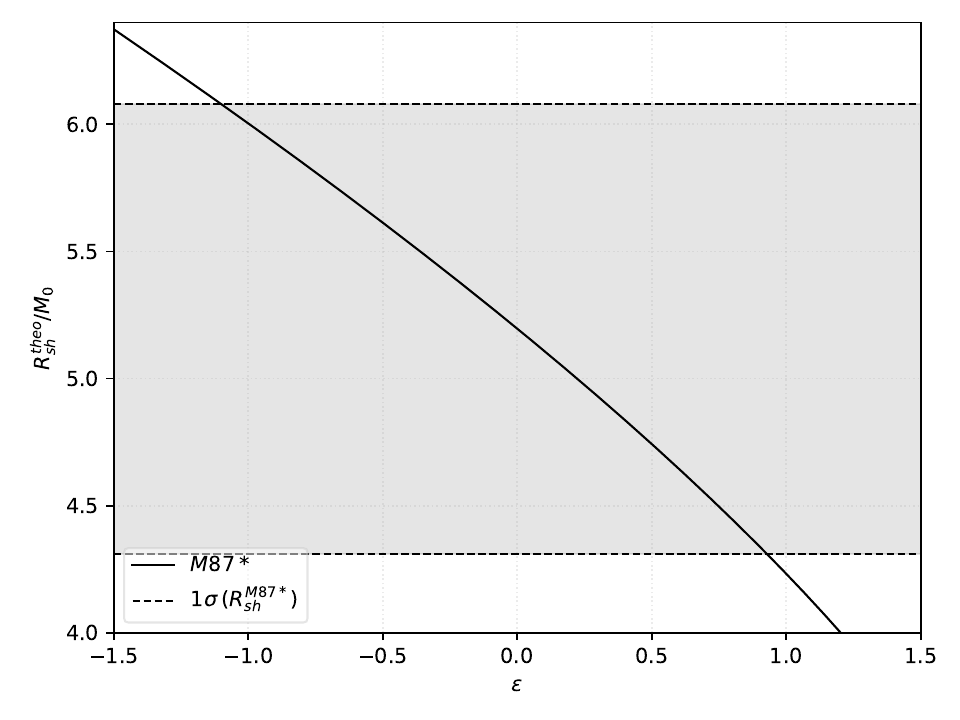}
    \caption{Constraining $\tilde{\epsilon}$. Left: Sgr. A*. Right: M87*. The shaded region is the allowable bounds based on $1\sigma$ level or $68\%$ confidence interval.}
    \label{shacons}
\end{figure*}

Using the above equation, we can study first the constraints to parameter $\tilde{\epsilon}$ using the data from EHT, which is summarized in Table \ref{tab1}. As for the allowed bounds at $68\%$ confidence level, we follow the bounds reported in papers \cite{EventHorizonTelescope:2019dse,EventHorizonTelescope:2022xnr,EventHorizonTelescope:2021dqv,EventHorizonTelescope:2021dqv,Vagnozzi:2022moj}. In particular, for Sgr. A* and M87*, these are $ 4.55M \leq R_{sh} \leq 5.22M$, and $ 4.31M \leq R_{sh} \leq 6.08M$, respectively. Our result is plotted in Fig. \ref{shacons} where the confidence intervals are included. Essentially, the plot shows how the shadow radius behaves if $\tilde{\epsilon}$ varies as we fix $x_\text{obs}$. The behavior of the curve is similar since we have used Eq. \eqref{eqshad}, but we can see that the upper and lower bounds for $\tilde{\epsilon}$ are different. These are listed in Table \ref{tab2}. Albeit the constraints to $\tilde{\epsilon}$ for Sgr. A* and M87* are different, the bounds are not that off if we compare. Interestingly, we see that there is a particular value for $\tilde{\epsilon}$ which perfectly describes the mean of the shadow radius. If one is used for $\tilde{\epsilon}$, it will introduce some small uncertainty of measurement on the other.
\begin{table}
    \centering
    \begin{tabular}{llll}
\hline
\hline
{} &   Mass($M_\odot$) & Angular diameter $\theta_\text{sh} = 2\alpha_\text{sh}$ ($\mu$as) &        Distance(kpc) \\
\hline
Sgr. A* &  $4.3 \pm 0.013$x$10^6$ (VLTI) &   $48.7 \pm 7$ (EHT) &  $8.277 \pm 0.033$ \\
M87*    & $6.5 \pm 0.90$x$10^9$ &   $42 \pm 3$ &  $16800$ \\
\hline
\end{tabular}
    \caption{Observational constraints of various black hole parameters based on EHT data.}
    \label{tab1}
\end{table}
\begin{table}[!ht]
    \centering
    \begin{tabular}{llll}
\hline
\hline
{} & $1\sigma$(upper/lower) \\
\hline
Sgr. A*  &      -0.03 / 0.706  \\
M87*     &      -1.110 / 0.92  \\
\hline
\end{tabular}
\caption{Values of $\tilde{\epsilon}$ based on the constraints imposed by the EHT data on the shadow radius.}
    \label{tab2}
\end{table}

\begin{figure*}
    \centering
    \includegraphics[width=0.48\textwidth]{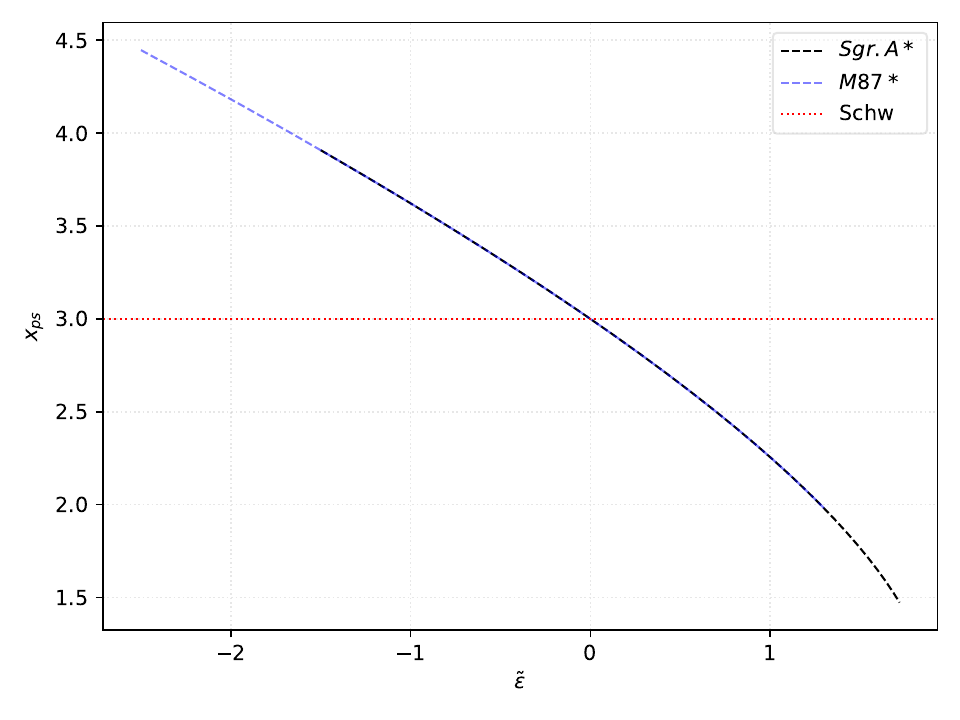}
    \includegraphics[width=0.48\textwidth]{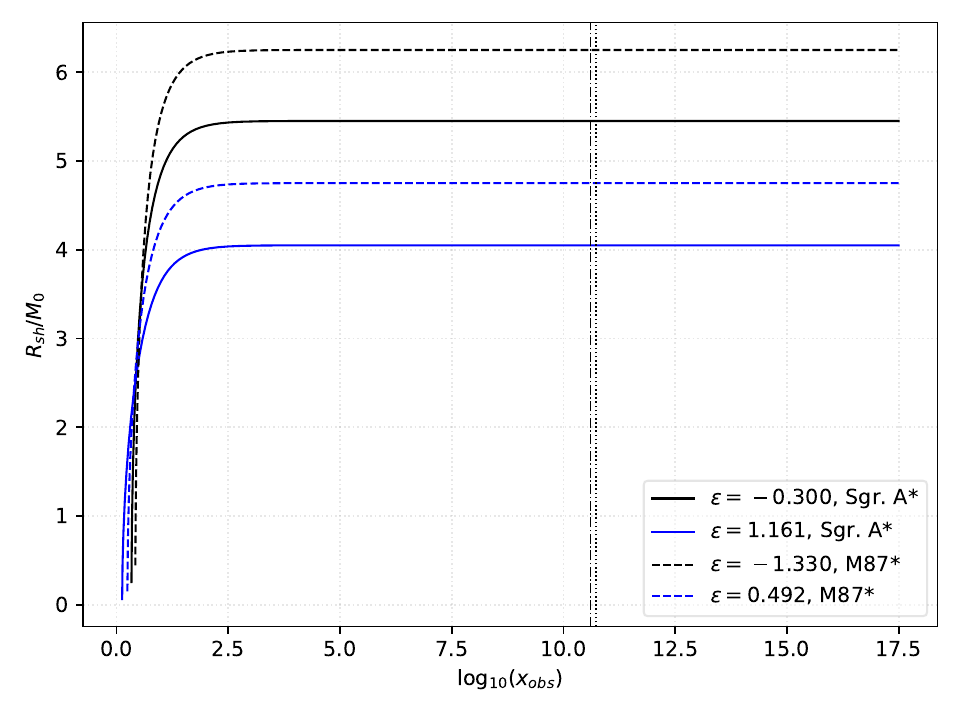}
    \caption{Left: Radius of the photon ring. The red horizontal dotted line corresponds to the photonsphere radius of a Schwarzschild BH. Right: Behavior of the shadow radius as $x_\text{obs}$ varies. The black vertical dotted and dashdot lines corresponds to our location from M87* and Sgr. A*, respectively.}
    \label{rps_sharad}
\end{figure*}

 Let us now use these bounds and pick some values for $\tilde{\epsilon}$ and examine the photonsphere behavior and the shadow radius behavior as $x_\text{obs}$ varies. The result is shown in Fig. \ref{rps_sharad}.
In the left panel of Fig. \ref{rps_sharad}, we can see how the photonsphere radius differs from the Schwarzschild case. Note that when $\tilde{\epsilon} = 0$, the photonsphere radius is merely the Schwarzschild case. In the right panel, it is shown how the shadow radius behaves as perceived by a static observer as $x_\text{obs}$ varies. We see how the shadow radius increases or decreases depending on the sign of $\tilde{\epsilon}$, but the overall trend of the curve follows the Schwarzschild case relative to our location from Sgr. A* and M87*. It implies that the metric under the effect of scale dependency is asymptotically flat since we saw no subtle deviation in the shadow behavior relative to the Schwarzschild case. In other words, the scale parameter $\tilde{\epsilon}$ merely shifts the shadow radius to a lower or a higher value while following the Schwarzschild trend.

\section{Weak deflection angle of photons and massive particles using Gauss-Bonnet Theorem } \label{sec4}

\begin{figure}[htp!]
   \centering
\includegraphics[scale=0.8]{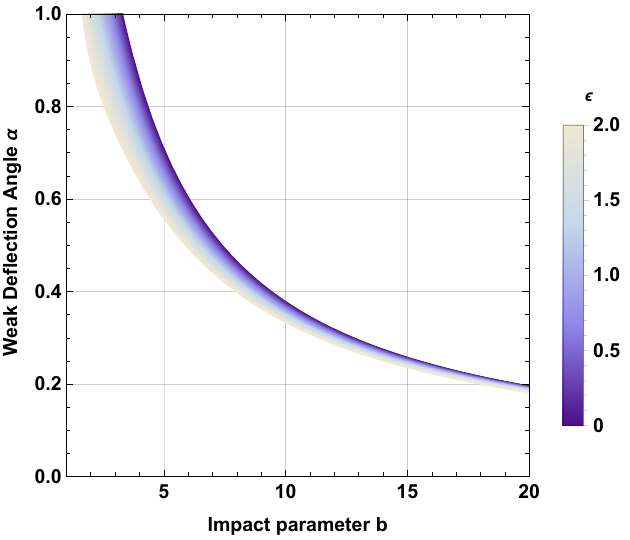}
        \caption{Depicting the relationship between the running parameter $ \epsilon $  and the weak deflection angle of a photon. }
    \label{fig:lensing33}
\end{figure}

In this section, we embark on a journey to explore the weak deflection angles experienced by photons and massive particles as they navigate around a RSDBH. To commence our investigation, we focus on photons, the messengers of light. By applying the null geodesics to the line element in Eq. (\ref{metricelement2}) for the equatorial plane $\theta=\pi/2$, we obtain the optical metric:
\begin{equation}
    \label{om}
    \mathrm{~d}t^2 = \frac{\mathrm{~d}r^2}{f(r)^2} + \frac{r^2}{f(r)} \mathrm{~d}\phi^2.
\end{equation}
Building upon the work of GBT (2008) \cite{Gibbons:2008rj}, we employ their boundary terms to calculate the weak deflection angle of a photon, which represents the slight bending of light as it passes around a massive object. The deflection angle, denoted by $\alpha$, is given by the integral: \cite{Gibbons:2008rj}
\begin{eqnarray}\label{int01}
\alpha=-\int\limits_{0}^{\pi}\int\limits_{{b}/{\sin \varphi}}^{\infty}\mathcal{K}\mathrm{~d}S,
\end{eqnarray}
where $\mathcal{K}=\frac{3 (r-2) \epsilon }{r^5}+\frac{3-2 r}{r^4}$.
For the specific optical metric defined in the previous section, the weak deflection angle of a photon simplifies to:
\begin{eqnarray}
\alpha=\frac{4}{\tilde{b}}-\frac{3 \pi }{4 \tilde{b}^2}-\frac{3 \pi  \tilde{\epsilon} }{4 \tilde{b}^2}.
\end{eqnarray}
where $\tilde{b} = b/M$ represents the distance of the closest approach between the photon and the black hole. Figure \ref{fig:lensing33} illustrates that increasing the value of $ \epsilon $ leads to a decrease in the weak deflection angle. This behavior is consistent with the expectation that stronger scale-dependent gravitational coupling should result in less bending of light.  

Subsequently, for computing the weak deflection angle of massive particles, we commence with a concise overview of the methodology \cite{Li:2020wvn}. We express a static, spherically symmetric spacetime in the following manner:
\begin{align}
    ds^2&=g_{\mu \nu}dx^{\mu}dx^{\nu} \nonumber \\
    &=-A(r)dt^2+B(r)dr^2+C(r)d\Omega^2,
\end{align}
then the Jacobi metric becomes
\begin{align}
    dl^2&=g_{ij}dx^{i}dx^{j} \nonumber \\
    &=[E^2-\mu^2A(r)]\left(\frac{B(r)}{A(r)}dr^2+\frac{C(r)}{A(r)}d\Omega^2\right),
\end{align}
where  $d\Omega^2=d\theta^2+r^2\sin^2\theta$, and $E$ is the energy of the massive particle
\begin{equation} \label{en}
    E = \frac{\mu}{\sqrt{1-v^2}},
\end{equation}
with particle velocity $v$.  The Jacobi metric in the equatorial plane can be written as
\begin{equation} \label{eJac}
    dl^2=(E^2-\mu^2A(r))\left(\frac{B(r)}{A(r)}dr^2+\frac{C(r)}{A(r)}d\phi^2\right),
\end{equation}
where the determinant of the above metric is $
    g=\frac{B(r)C(r)}{A(r)^2}(E^2-\mu^2 A(r))^2.$
    Using the method defined in \cite{Li:2020wvn}, the weak deflection angle is calculated by using
\begin{equation} \label{eIshi}
    \hat{\alpha} = \iint_{D}K\sqrt{g}drd\phi + \phi_{\text{RS}},
\end{equation}
where $r_\text{co}$ is the radius of the particle's circular orbit with the radial position of the source $S$ and receiver $R$. 
Moreover, $\phi_\text{RS}$ is the angle of coordinate position between the source/ receiver $\phi_\text{RS} = \phi_\text{R}-\phi_\text{S}$, calculated by iterative method as:
\begin{align} \label{fu}
    F(u) = \left(\frac{du}{d\phi}\right)^2= \frac{C(u)^2u^4}{A(u)B(u)}\Bigg[\left(\frac{E}{J}\right)^2-A(u)\left(\frac{1}{J^2}+\frac{1}{C(u)}\right)\Bigg],
\end{align}
where $r = 1/u$ and the angular momentum of massive particle with impact parameter $b$: $J = \frac{\mu v b}{\sqrt{1-v^2}}$.
With Eq. \eqref{fu}, we find
\begin{align}
    F(u) = \frac{E^2-1}{J^2}-u^2-u^2\left(\frac{1}{J^2}+u^2\right)\epsilon M^3
    + \left(\frac{1}{J^2}+u^2\right)2Mu.
\end{align}
Iteration method provides us this solution:
\begin{equation} \label{orb}
    u(\phi) = \frac{\sin(\phi)}{b}+\frac{1+v^2\cos^2(\phi)}{b^2v^2}M.
\end{equation}
Then we have the relation of
\begin{equation} \label{gct}
    \int_{r_\text{co}}^{r(\phi)} K\sqrt{g}dr = -\frac{A(r)\left(E^{2}-A(r)\right)C'-E^{2}C(r)A(r)'}{2A(r)\left(E^{2}-A(r)\right)\sqrt{B(r)C(r)}}\bigg|_{r = r(\phi)},
\end{equation}
where $K=-\frac{1}{\sqrt{g}}\left[\frac{\partial}{\partial r}\left(\frac{\sqrt{g}}{g_{rr}}\Gamma_{r\phi}^{\phi}\right)\right]$. Note that
$ \left[\int K\sqrt{g}dr\right]\bigg|_{r=r_\text{co}} = 0.$

We then calculate the weak deflection angle as follows,
\begin{align} \label{ewda}
    \hat{\alpha} = \int^{\phi_\text{R}}_{\phi_\text{S}} \left[-\frac{A(r)\left(E^{2}-A(r)\right)C'-E^{2}C(r)A(r)'}{2A(r)\left(E^{2}-A(r)\right)\sqrt{B(r)C(r)}}\bigg|_{r = r(\phi)}\right] d\phi + \phi_\text{RS}.
\end{align}
Using Eq. \eqref{orb} in Eq. \eqref{gct}, one can see that
\begin{align} \label{gct2}
    \left[\int K\sqrt{g}dr\right]\bigg|_{r=r_\phi} &= -\frac{\left(2E^{2}-1\right)M(\cos(\phi_\text{R})-\cos(\phi_\text{S}))}{\left(E^{2}-1\right)b} + \frac{M^{2} \phi_\text{RS}  \left(7 E^{4} v^{2}+8 E^{4}-14 E^{2} v^{2}-12 E^{2}+3 v^{2}+4\right)}{4 b^{2} v^{2} \left(E^{2}-1\right)^{2}} \\ \nonumber
    &-\frac{\left(3E^{2}-1\right)\epsilon M^3\left[\phi_\text{RS}-\frac{(\sin(2\phi_\text{R})-\sin(2\phi_\text{S})}{2}\right]}{4\left(E^{2}-1\right)b^{2}}.
\end{align}
Then using the Eq. \eqref{orb}, we obtain the $\phi$ for the source and receiver, respectively
\begin{align} \label{s}
    \phi_\text{S} =\arcsin(bu)+\frac{M\left[v^{2}\left(b^{2}u^{2}-1\right]-1\right)}{bv^{2}\sqrt{1-b^{2}u^{2}}}
    +\frac{\epsilon M^3}{2b^{2}v^{2}\sqrt{1-b^{2}u^{2}}},
\end{align}
\begin{align} \label{r}
    \phi_\text{R} =\pi -\arcsin(bu)-\frac{M\left[v^{2}\left(b^{2}u^{2}-1\right]-1\right)}{bv^{2}\sqrt{1-b^{2}u^{2}}}
    -\frac{\epsilon M^3}{2b^{2}v^{2}\sqrt{1-b^{2}u^{2}}},
\end{align}
hence, we wrote $\phi_\text{RS} = \pi - 2\phi_\text{S}$. With the help of $ \cos(\pi-\phi_\text{S})=-\cos(\phi_\text{S}) \text{and}, 
    \cot(\pi-\phi_\text{S})=-\cot(\phi_\text{S}),$ we calculate $\cos(\phi_\text{S})$ as
\begin{align} \label{cs}
    \cos(\phi_\text{S}) = \sqrt{1-b^{2}u^{2}}-\frac{Mu\left[v^{2}\left(b^{2}u^{2}-1\right)-1\right]}{\sqrt{v^{2}\left(1-b^{2}u^{2}\right)}}
    -\frac{\epsilon M^3u}{\sqrt{2}\sqrt{bv^{2}\left(1-b^{2}u^{2}\right)}},
\end{align}
and $\cot(\phi_\text{S})$ as
\begin{align} \label{cr}
    \cot(\phi_\text{S}) = \frac{\sqrt{1-b^{2}u^{2}}}{bu}+\frac{M\left[v^{2}(-b^{2}u^{2}+1)+1\right]}{b^{3}u^{2}v^{2}\sqrt{1-b^{2}u^{2}}}-\frac{\epsilon M^3}{2b^{4}u^{2}v^{2}\sqrt{1-b^{2}u^{2}}}.
\end{align}
By substituting Eqs. \eqref{s}-\eqref{cr} into the Eq. \eqref{ewda}, obtain 
\begin{align} \label{ewda_exact}
    \hat{\alpha} &\sim \frac{M\left(v^{2}+1\right)}{bv^{2}}\left(\sqrt{1-b^{2}u_\text{R}^{2}}+\sqrt{1-b^{2}u_\text{S}^{2}}\right) + \frac{3(v^2 + 4)M^2}{4b^{2}v^{2}}\left[\pi-(\arcsin(bu_\text{R})+\arcsin(bu_\text{S}))\right]
    \\ \nonumber
    &-\frac{\epsilon M^3\left(v^{2}+2\right)}{4b^{2}v^{2}}\left[\pi-(\arcsin(bu_\text{R})+\arcsin(bu_\text{S}))\right],
\end{align}
where $u_\text{S}$ and $u_\text{R}$ are for the finite distances. By assuming $b^2u^2 \sim 0$ and we find
\begin{align} \label{ewda_approx}
    \hat{\alpha} \sim \frac{2M\left(v^{2}+1\right)}{bv^{2}} + \frac{3\pi(v^2 + 4)M^2}{4b^{2}v^{2}} -\frac{\epsilon M^3 \pi\left(v^{2}+2\right)}{4b^{2}v^{2}}.
\end{align}
For the null particles ($v = 1$), the above result reduces to
\begin{align} \label{wdafin}
    \hat{\alpha} \sim \frac{4M}{b} + \frac{15\pi M^2}{4b^{2}}-\frac{3\pi\epsilon M^3 }{4b^{2}}.
\end{align}

\begin{figure}
   \centering
    \includegraphics[width=0.48\textwidth]{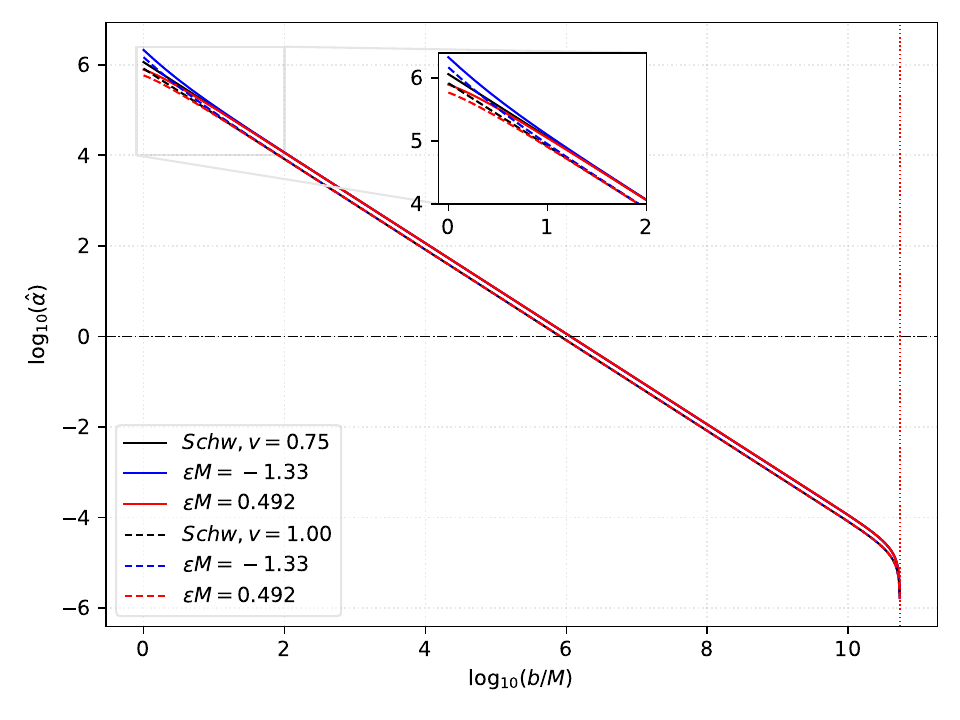}    
    \caption{Figure shows the weak deflection angle $\alpha$ versus impact parameter $b/M$ for different values of $\epsilon M$. To illustrate the deviation, we used the log-log plot.}
    \label{fig:lensing3}
\end{figure}

Hence, one can see that $\hat{\alpha}$ will change due to the substitution $\tilde{b} = b/M$, and $\tilde{\epsilon} = \epsilon M$, which is just a matter of convenience. Furthermore, since finite-distance correction was introduced, we expect more sensitivity in precision for higher orders in $M$. Nonetheless, this result aligns with the deflection angle calculated in Ref. \cite{Scardigli:2014qka}. We plot Eq. \eqref{ewda_exact} as shown in Fig. \ref{fig:lensing3}, where we used M87* and its actual distance from us, as well as the values for $\tilde{\epsilon}$ found between the bounds in Table \ref{tab2}. 
Fig. \ref{fig:lensing3} shows how $\hat{\alpha}$ varies as $b/M$ changes while the observer is at $r_\text{obs}$. First, massive particle deflection gives a larger value than photon deflection. The trend, however, remains the same wherein $\hat{\alpha}$ increases as $b/M$ decreases. The effect of $\epsilon$ is also evident as $b/M \rightarrow 0$ and immediately weakens. Thus, upon considering the finite distance, one could only observe the deviation due to $\epsilon$ as photons grazes slightly near the critical impact parameter.

\section{Rigorous Bounds of Greybody factors} \label{sec5}
\subsection{For scalar fields emitted by a black hole}
In this section, we turn to analyse of greybody factor bounds for a regular scale-dependent black hole. We follow the method defined in \cite{Visser:1998ke,Boonserm:2008zg,Boonserm:2017qcq,Okyay:2021nnh,Boonserm:2019mon,Liu:2022ygf,Yang:2022ifo,Gray:2015xig,Boonserm:2014fja,Boonserm:2014rma,Boonserm:2013dua,Ngampitipan:2012dq,Boonserm:2009zba,Boonserm:2009mi,Lambiase:2023zeo,Gogoi:2023fow,Ahmed:2016lou,Al-Badawi:2022aby,Al-Badawi:2021wdm,Javed:2021ymu,Javed:2022kzf,Javed:2022rrs,Mangut:2023oxa,Sakalli:2023pgn}. We begin by examining the Klein-Gordon equation for a massless scalar field:
\begin{equation}
\frac{1}{\sqrt{-g}}\partial_{\mu}\left(\sqrt{-g}g^{\mu\nu}\partial_{\nu}\Phi\right) = 0.
\end{equation}

The solutions of the wave equation in spherical coordinates take the simplest form when they are expressed in terms of the spherical Bessel functions. These functions are solutions to the Helmholtz equation, which is a special case of the wave equation. The spherical Bessel functions are labeled by two integers, $\ell$ and $m$, where $\ell$ is the non-negative orbital quantum number and $m$ is the magnetic quantum number. The general form of a solution to the wave equation in spherical coordinates is given by:
\begin{equation}
\Phi(t, r, \Omega) = e^{i\omega t}\frac{\psi(r)}{r}Y_{\ell m}(\Omega),
\end{equation}
being the spherical harmonics $Y_{\ell m}(\Omega)$.
Taking the latter into account, the Klein-Gordon equation reduces to this form:
\begin{eqnarray}
\frac{\omega^{2}r^{2}}{f(r)} + \frac{r}{\psi(r)}\frac{d}{dr}\left[r^{2}f(r)\frac{d}{dr}\left(\frac{\psi(r)}{r}\right)\right] 
+ \frac{1}{Y(\Omega)}\left[\frac{1}{\sin\theta}\frac{\partial}{\partial\theta}\left(\sin\theta\frac{\partial Y(\Omega)}{\partial\theta}\right)\right] &&\nonumber\\
+ \frac{1}{\sin^{2}\theta}\frac{1}{Y(\Omega)}\frac{\partial^{2}Y(\Omega)}{\partial\phi^{2}} &=& 0,\label{KG}
\end{eqnarray}
and the angular part is described by
\begin{equation}
\frac{1}{\sin\theta}\frac{\partial}{\partial\theta}\left(\sin\theta\frac{\partial Y(\Omega)}{\partial\theta}\right) + \frac{1}{\sin^{2}\theta}\frac{\partial^{2}Y(\Omega)}{\partial\phi^{2}} = -\ell(\ell + 1)Y(\Omega).
\end{equation}
The Klein-Gordon equation (Eq. (\ref{KG})) is left with the radial part in the tortoise coordinate $r_{*}$:
\begin{equation}
\frac{d^{2}\psi(r)}{dr_{*}^{2}} + \left[\omega^{2} - V(r)\right]\psi(r) = 0.
\end{equation}
Please, be aware and notice we have used the definition of the tortoise coordinate, given by
\begin{align}
    r_{*} = \int f(r)^{-1} \mathrm{d}r
\end{align}
Taking advantage of the last definition, the effective potential $V(r)$ takes the form
\begin{equation}
V(r) = f(r) 
\Bigg[ 
\frac{\ell(\ell + 1)}{r^{2}} + \frac{f'(r)}{r} 
\Bigg].
\label{poten}
\end{equation}

Utilizing the effective potential, $V(r)$, we delve into the analysis of the lower rigorous bound for the greybody factor of the four-dimensional regular scale-dependent black hole to investigate the influence of $\epsilon$ on the bound \cite{Visser:1998ke,Boonserm:2008zg}:
\begin{equation} \label{bound}
T \geq \operatorname{sech}^{2}\left(\frac{1}{2 \omega} \int_{-\infty}^{\infty}\left|V(r)\right| \frac{d r}{f(r)} \right),
\end{equation}
When the cosmological constant is incorporated, the boundary of the aforementioned formula undergoes slight modifications, as outlined in \cite{Boonserm:2019mon}:
\begin{equation}
T \geq T_{b}=\operatorname{sech}^{2}\left(\frac{1}{2 \omega} \int_{r_{H}}^{R_{H}} \frac{|V(r)|}{f(r)} d r\right)=\operatorname{sech}^{2}\left(\frac{A_{\ell}}{2 \omega}\right),
\end{equation}
where the factor $A_{\ell}$ is defined according to the following expression:
\begin{equation}
A_{\ell}=\int_{r_{H}}^{R_{H}} \frac{|V(r)|}{f(r)} d r=\int_{r_{H}}^{R_{H}}\left|\frac{\ell(\ell+1)}{r^{2}}+\frac{f(r)^{\prime}}{r}\right| d r.
\end{equation}
Hence, the bounds of the greybody factor of bosons are given by
\begin{equation}T \geq T_{b}=\text{sech}^2\left(\frac{-\frac{l^2}{R_H}-\frac{l}{R_H}+\frac{36 \left(\epsilon -6 R_H\right)}{\left(6 R_H+\epsilon \right){}^3}+\frac{l^2}{r_{\text{h}}}+\frac{l}{r_{\text{h}}}-\frac{36 \left(\epsilon -6 r_{\text{h}}\right)}{\left(6 r_{\text{h}}+\epsilon \right){}^3}}{2 \omega }\right).
\end{equation}

\begin{figure*}[!ht]
    \centering
    \includegraphics[width=0.68\textwidth]{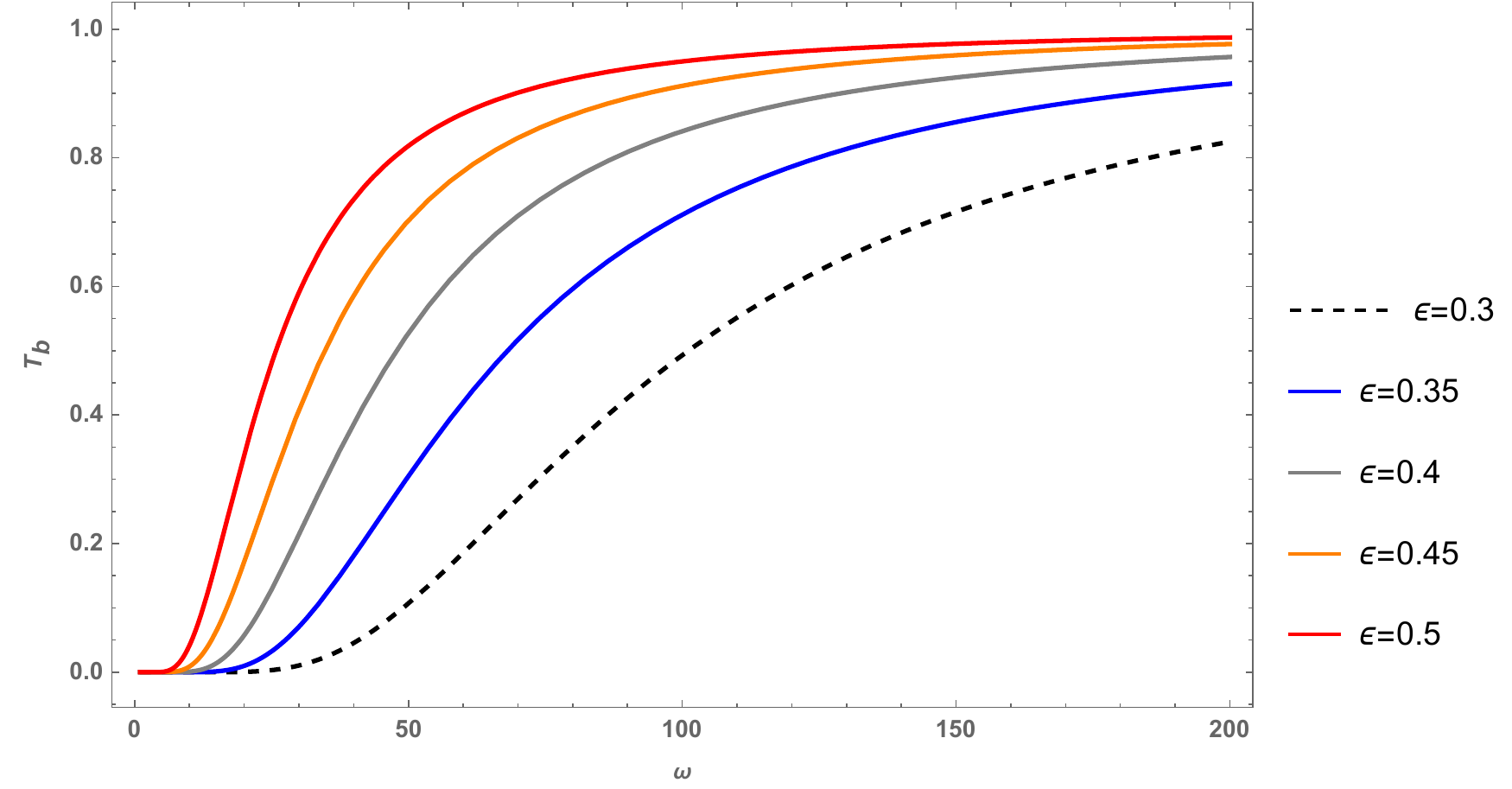}
    \includegraphics[width=0.68\textwidth]{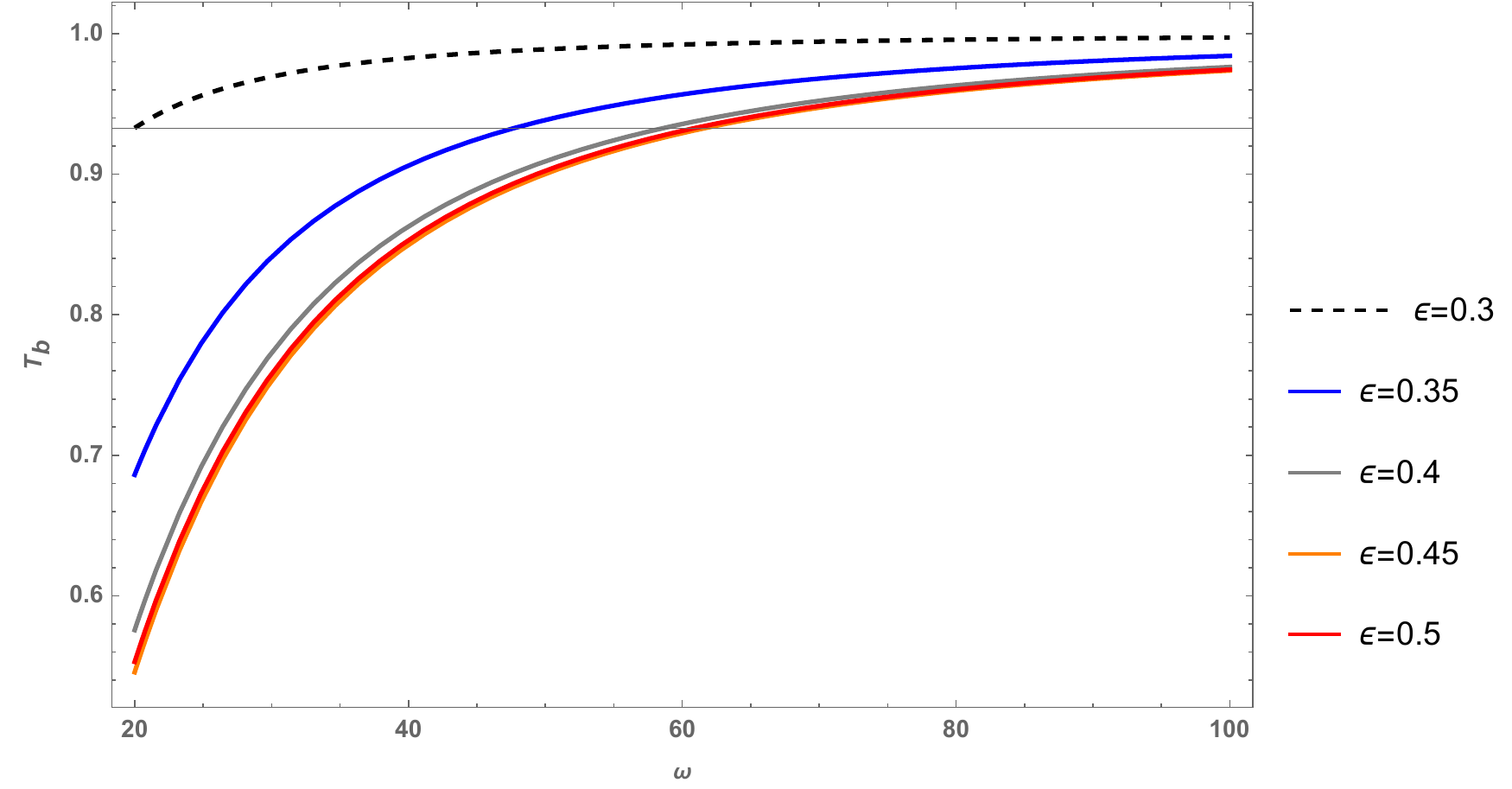}
    \caption{Left: The greybody bounding of bosons is plotted for $\ell=0$, $M_0=1$ and various values of $\epsilon$  . Right:  The greybody bounding is plotted for $\ell=1$ and various values of $\epsilon$  .}
    \label{fig:greybody}
\end{figure*}

To illustrate the behavior of the bound, we numerically calculate and plot it in Figure \ref{fig:greybody} for both $\ell=0$  and $\ell=1$. The graph reveals that as the parameter $\epsilon$  increases, the bound of the greybody factor for bosons also increases. This observation suggests that RSDBHs exhibit favorable barrier properties.

\subsection{For photons emitted by a black hole}
Additionally, electromagnetic perturbations are governed by Maxwell's equations, which are expressed as:
\begin{equation}
F_{; \nu}^{\mu \nu}=0, \quad F_{\mu \nu} = \partial_{\mu} A_{\nu}-\partial_{\nu} A_{\mu},
\end{equation} where $(A_{\mu})$ represents the Maxwell potential, $(F_{\mu \nu})$ denotes the electromagnetic field strength, and a semi-colon indicates covariant differentiation.
\begin{equation}
V_{E M}(r)=f(r)\left(\frac{l(l+1)}{r^{2}}\right), \quad l \geq 1
\end{equation}

Consequently, the bounds for the greybody factor of photons are expressed as:
\begin{equation}
T \geq T_{b}=\operatorname{sech}^{2}\left(\frac{1}{2 \omega} \int_{-\infty}^{\infty}|V| \frac{d r}{f(r)}\right),
\end{equation}
and
\begin{equation}
T \geq T_{b}=\operatorname{sech}^2\left(\frac{\frac{l (l+1)}{r_{\text{outer}}}-\frac{l (l+1)}{r_H}}{2 \omega }\right).
\end{equation}

\begin{figure*}[!ht]
    \centering
    \includegraphics[width=0.68\textwidth]{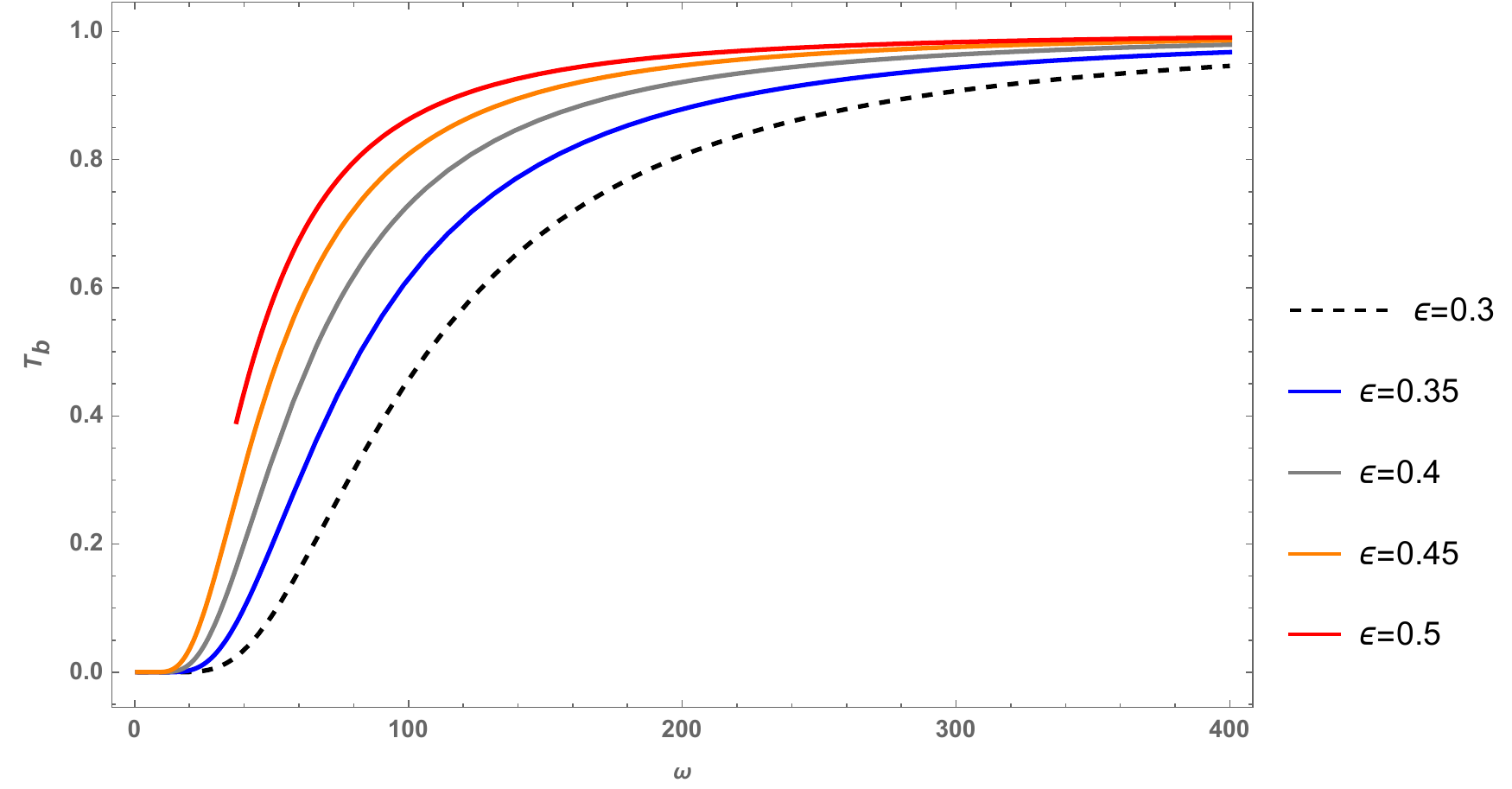}
    \caption{The greybody bounding of photons is plotted for $\ell=1$ and various values of $\epsilon$  }
    \label{fig:greybody2}
\end{figure*}

The bound for the greybody factor of photons can be numerically determined and illustrated in Figure \ref{fig:greybody2}, with one example showcasing the case for $\ell=1$, $M_0=1$, and various values of $\epsilon$. As the value of $\epsilon$ increases, the bound for the greybody factor of photons also increases. This observation suggests that RSDBHs exhibit favorable barrier properties.

\section{Conclusion} \label{concl}
Given the high level of interest in the study of black hole shadow, lensing, and greybody factor, this paper focuses on these physical phenomena in the context of RSDBHs in order to establish constraints on the scale-dependent parameter.

Our first goal in this paper to find constraints to the scale-dependent parameter $\tilde{\epsilon}$ (unit of inverse of length) associated with the gravitational coupling. Results using the EHT data indicates that there are different values of $\tilde{\epsilon}$ coinciding with the reported black hole shadow radius. For instance, we found a positive value for Sgr. A* ($\tilde{\epsilon} = 0.492$) and a negative value for M87* ($\tilde{\epsilon}=-0.361$). The bounds were also obtained up to $2\sigma$ levels, indicating that any fluctuation of the shadow radius from the mean could be the effect of the scale-dependent parameter. We used these bounds to analyze the resulting photonsphere radius and shadow radius perceived by an observer at $x_\text{obs}$. While the behavior of the shadow radius curve is reminiscent of the Schwarzschild case, it increases or decreases depending on the value of $\tilde{\epsilon}$. The shadow radius is only smaller near the black hole, but fluctuations of its value for a given scale-dependent parameter is not observer even at vast distances (see Fig. \ref{rps_sharad}).

 We have also analyzed the weak deflection angle behavior using the constraints to the supermassive black hole M87*. Result showed that the effect of $\epsilon$ depends on the impact parameter, where the effect is only seen when $b/M$ is slightly near the critical impact parameter. It is then very hard to discern the deviation when the impact parameter becomes very large.

 Lastly, we examined the greybody bounds for bosons and photons emitted by RSDBHs. It was found that as the value of $\epsilon$ increases, the bounds for the greybody factor increases for both bosons and photons.

Regarding the results of this study, studies can be undertaken to further explore the effect of the scale-dependent parameter. For instance, since the effect weak deflection angle immediately vanishes at large impact parameters, one may consider the strong deflection angle. The effect of $\epsilon$ could be strong in such domain, since the strong deflection angle regime is almost at the critical impact parameter.

Finally, our results provide a potential direction for research on black hole models where quantum features can be incorporated. In particular, our work establishes constraints on the scale-dependent parameter, which plays a relevant role since it controls the strength of the quantum effects on the classical background. A future work in this direction is the study of optical properties with more realistic features, such as the study of shadows and rings of this scale-dependent black hole with thin spherical emission.
The impact of this and subsequent related work becomes evident because this work allows us to falsify if quantum-like black holes are consistent with observations. According to our results, this solution is consistent with observations for some concrete values of the scale-dependent parameter $\epsilon$.

\section{ACKNOWLEDGMENTS}
We would like to thank the referees for reading the manuscript and for their useful suggestions to improve the quality of this work.
A. R. acknowledges financial support from the Generalitat Valenciana through PROMETEO PROJECT CIPROM/2022/13.
A. R. is funded by the María Zambrano contract ZAMBRANO 21-25 (Spain) (with funding from NextGenerationEU).  
A. {\"O}. and R. P. would like to acknowledge networking support by the COST Action CA18108 - Quantum gravity phenomenology in the multi-messenger approach (QG-MM). A. {\"O}. and R. P. also would like to acknowledge networking support by the COST Action CA21106 - COSMIC WISPers in the Dark Universe: Theory, astrophysics and experiments (CosmicWISPers), the COST Action CA22113 - Fundamental challenges in theoretical physics (THEORY-CHALLENGES), and the COST Action CA21136 - Addressing observational tensions in cosmology with systematics and fundamental physics: (FUNDAMENTAL PHYSICS).


\begin{thebibliography}{0}%
\makeatletter
\providecommand \@ifxundefined [1]{%
 \@ifx{#1\undefined}
}%
\providecommand \@ifnum [1]{%
 \ifnum #1\expandafter \@firstoftwo
 \else \expandafter \@secondoftwo
 \fi
}%
\providecommand \@ifx [1]{%
 \ifx #1\expandafter \@firstoftwo
 \else \expandafter \@secondoftwo
 \fi
}%
\providecommand \natexlab [1]{#1}%
\providecommand \enquote  [1]{``#1''}%
\providecommand \bibnamefont  [1]{#1}%
\providecommand \bibfnamefont [1]{#1}%
\providecommand \citenamefont [1]{#1}%
\providecommand \href@noop [0]{\@secondoftwo}%
\providecommand \href [0]{\begingroup \@sanitize@url \@href}%
\providecommand \@href[1]{\@@startlink{#1}\@@href}%
\providecommand \@@href[1]{\endgroup#1\@@endlink}%
\providecommand \@sanitize@url [0]{\catcode `\\12\catcode `\$12\catcode
  `\&12\catcode `\#12\catcode `\^12\catcode `\_12\catcode `\%12\relax}%
\providecommand \@@startlink[1]{}%
\providecommand \@@endlink[0]{}%
\providecommand \url  [0]{\begingroup\@sanitize@url \@url }%
\providecommand \@url [1]{\endgroup\@href {#1}{\urlprefix }}%
\providecommand \urlprefix  [0]{URL }%
\providecommand \Eprint [0]{\href }%
\providecommand \doibase [0]{http://dx.doi.org/}%
\providecommand \selectlanguage [0]{\@gobble}%
\providecommand \bibinfo  [0]{\@secondoftwo}%
\providecommand \bibfield  [0]{\@secondoftwo}%
\providecommand \translation [1]{[#1]}%
\providecommand \BibitemOpen [0]{}%
\providecommand \bibitemStop [0]{}%
\providecommand \bibitemNoStop [0]{.\EOS\space}%
\providecommand \EOS [0]{\spacefactor3000\relax}%
\providecommand \BibitemShut  [1]{\csname bibitem#1\endcsname}%
\let\auto@bib@innerbib\@empty
\end{thebibliography}%


\begin{thebibliography}{999}
\bibliographystyle{apsrev}

\bibitem{LIGOScientific:2016wkq}
T.~D.~Abbott \textit{et al.} [LIGO Scientific and Virgo],
Phys. Rev. X \textbf{6}, no.4, 041014 (2016)
doi:10.1103/PhysRevX.6.041014
[arXiv:1606.01210 [gr-qc]].

\bibitem{LIGOScientific:2016vlm}
B.~P.~Abbott \textit{et al.} [LIGO Scientific and Virgo],
Phys. Rev. Lett. \textbf{116}, no.24, 241102 (2016)
doi:10.1103/PhysRevLett.116.241102
[arXiv:1602.03840 [gr-qc]].

\bibitem{Schwarzschild:1916uq}
K.~Schwarzschild,
Sitzungsber. Preuss. Akad. Wiss. Berlin (Math. Phys. ) \textbf{1916}, 189-196 (1916)
[arXiv:physics/9905030 [physics]].

\bibitem{Hawking:1974rv}
S.~W.~Hawking,
Nature \textbf{248}, 30-31 (1974)
doi:10.1038/248030a0

\bibitem{Hawking:1975vcx}
S.~W.~Hawking,
Commun. Math. Phys. \textbf{43}, 199-220 (1975)
[erratum: Commun. Math. Phys. \textbf{46}, 206 (1976)]
doi:10.1007/BF02345020

\bibitem{Hawking:1973uf}
S.~W.~Hawking and G.~F.~R.~Ellis,
Cambridge University Press, 2011,
ISBN 978-0-521-20016-5, 978-0-521-09906-6, 978-0-511-82630-6, 978-0-521-09906-6
doi:10.1017/CBO9780511524646

\bibitem{Penrose:1964wq}
R.~Penrose,
Phys. Rev. Lett. \textbf{14}, 57-59 (1965)
doi:10.1103/PhysRevLett.14.57

\bibitem{Hawking:1965mf}
S.~Hawking,
Phys. Rev. Lett. \textbf{15}, 689-690 (1965)
doi:10.1103/PhysRevLett.15.689

\bibitem{Hawking:1966sx}
S.~Hawking,
Proc. Roy. Soc. Lond. A \textbf{294}, 511-521 (1966)
doi:10.1098/rspa.1966.0221

\bibitem{Hawking:1966jv}
S.~Hawking,
Proc. Roy. Soc. Lond. A \textbf{295}, 490-493 (1966)
doi:10.1098/rspa.1966.0255

\bibitem{Hawking:1967ju}
S.~Hawking,
Proc. Roy. Soc. Lond. A \textbf{300}, 187-201 (1967)
doi:10.1098/rspa.1967.0164

\bibitem{Borde:1994ai}
A.~Borde,
Phys. Rev. D \textbf{50}, 3692-3702 (1994)
doi:10.1103/PhysRevD.50.3692
[arXiv:gr-qc/9403049 [gr-qc]].

\bibitem{Hawking:1970zqf}
S.~W.~Hawking and R.~Penrose,
Proc. Roy. Soc. Lond. A \textbf{314}, 529-548 (1970)
doi:10.1098/rspa.1970.0021

\bibitem{Clarke:1975ph}
C.~J.~S.~Clarke,
Commun. Math. Phys. \textbf{41}, 65-78 (1975)
doi:10.1007/BF01608548

\bibitem{Israel:1967za}
W.~Israel,
Commun. Math. Phys. \textbf{8}, 245-260 (1968)
doi:10.1007/BF01645859

\bibitem{Dymnikova:1992ux}
I.~Dymnikova,
Gen. Rel. Grav. \textbf{24}, 235-242 (1992)
doi:10.1007/BF00760226

\bibitem{Nicolini:2005vd}
P.~Nicolini, A.~Smailagic and E.~Spallucci,
Phys. Lett. B \textbf{632}, 547-551 (2006)
doi:10.1016/j.physletb.2005.11.004
[arXiv:gr-qc/0510112 [gr-qc]].

\bibitem{Hayward:2005gi}
S.~A.~Hayward,
Phys. Rev. Lett. \textbf{96}, 031103 (2006)
doi:10.1103/PhysRevLett.96.031103
[arXiv:gr-qc/0506126 [gr-qc]].

\bibitem{Xiang:2013sza}
L.~Xiang, Y.~Ling and Y.~G.~Shen,
Int. J. Mod. Phys. D \textbf{22}, 1342016 (2013)
doi:10.1142/S0218271813420169
[arXiv:1305.3851 [gr-qc]].

\bibitem{Ghosh:2014pba}
S.~G.~Ghosh,
Eur. Phys. J. C \textbf{75}, no.11, 532 (2015)
doi:10.1140/epjc/s10052-015-3740-y
[arXiv:1408.5668 [gr-qc]].

\bibitem{Frolov:2016pav}
V.~P.~Frolov,
Phys. Rev. D \textbf{94}, no.10, 104056 (2016)
doi:10.1103/PhysRevD.94.104056
[arXiv:1609.01758 [gr-qc]].

\bibitem{Frolov:2017rjz}
V.~P.~Frolov and A.~Zelnikov,
Phys. Rev. D \textbf{95}, no.12, 124028 (2017)
doi:10.1103/PhysRevD.95.124028
[arXiv:1704.03043 [hep-th]].

\bibitem{Bardeen:1968}
J. M. Bardeen
``Non-singular general-relativistic gravitational collapse,''
in Proceedings of GR5, Tbilisi, U.S.S.R, p.174 (1968).

\bibitem{Ayon-Beato:2000mjt}
E.~Ayon-Beato and A.~Garcia,
Phys. Lett. B \textbf{493}, 149-152 (2000)
doi:10.1016/S0370-2693(00)01125-4
[arXiv:gr-qc/0009077 [gr-qc]].

\bibitem{Rodrigues:2018bdc}
M.~E.~Rodrigues and M.~V.~de Sousa Silva,
JCAP \textbf{06}, 025 (2018)
doi:10.1088/1475-7516/2018/06/025
[arXiv:1802.05095 [gr-qc]].

\bibitem{Balart:2014jia}
L.~Balart and E.~C.~Vagenas,
Phys. Lett. B \textbf{730}, 14-17 (2014)
doi:10.1016/j.physletb.2014.01.024
[arXiv:1401.2136 [gr-qc]].

\bibitem{Bronnikov:2017sgg}
K.~A.~Bronnikov,
Int. J. Mod. Phys. D \textbf{27}, no.06, 1841005 (2018)
doi:10.1142/S0218271818410055
[arXiv:1711.00087 [gr-qc]].

\bibitem{He:2017ujy}
Y.~He and M.~S.~Ma,
Phys. Lett. B \textbf{774}, 229-234 (2017)
doi:10.1016/j.physletb.2017.09.044
[arXiv:1709.09473 [gr-qc]].

\bibitem{1950SRToh..34..160N} 
Nariai, H.\ 1950, Sci. Rep. Tohoku Univ. Eighth Ser., 34, 160


\bibitem{Balart:2014cga}
L.~Balart and E.~C.~Vagenas,
Phys. Rev. D \textbf{90}, no.12, 124045 (2014)
doi:10.1103/PhysRevD.90.124045
[arXiv:1408.0306 [gr-qc]].

\bibitem{Synge:1966okc}
J.~L.~Synge,
Mon. Not. Roy. Astron. Soc. \textbf{131}, no.3, 463-466 (1966)
doi:10.1093/mnras/131.3.463

\bibitem{Luminet:1979nyg}
J.~P.~Luminet,
Astron. Astrophys. \textbf{75}, 228-235 (1979)

\bibitem{EventHorizonTelescope:2019dse}
K.~Akiyama \textit{et al.} [Event Horizon Telescope],
Astrophys. J. Lett. \textbf{875}, L1 (2019)
doi:10.3847/2041-8213/ab0ec7
[arXiv:1906.11238 [astro-ph.GA]].

\bibitem{EventHorizonTelescope:2022xnr}
K.~Akiyama \textit{et al.} [Event Horizon Telescope],
Astrophys. J. Lett. \textbf{930}, no.2, L12 (2022)
doi:10.3847/2041-8213/ac6674

\bibitem{Dokuchaev:2019jqq}
V.~I.~Dokuchaev and N.~O.~Nazarova,
Usp. Fiz. Nauk \textbf{190}, no.6, 627-647 (2020)
doi:10.3367/UFNe.2020.01.038717
[arXiv:1911.07695 [gr-qc]].

\bibitem{Perlick:2015vta}
V.~Perlick, O.~Y.~Tsupko and G.~S.~Bisnovatyi-Kogan,
Phys. Rev. D \textbf{92}, no.10, 104031 (2015)
doi:10.1103/PhysRevD.92.104031
[arXiv:1507.04217 [gr-qc]].

\bibitem{Perlick:2021aok}
V.~Perlick and O.~Y.~Tsupko,
Phys. Rept. \textbf{947}, 1-39 (2022)
doi:10.1016/j.physrep.2021.10.004
[arXiv:2105.07101 [gr-qc]].

\bibitem{Afrin:2021imp}
M.~Afrin, R.~Kumar and S.~G.~Ghosh,
Mon. Not. Roy. Astron. Soc. \textbf{504}, 5927-5940 (2021)
doi:10.1093/mnras/stab1260
[arXiv:2103.11417 [gr-qc]].

\bibitem{Atamurotov:2015nra}
F.~Atamurotov and B.~Ahmedov,
Phys. Rev. D \textbf{92}, 084005 (2015)
doi:10.1103/PhysRevD.92.084005
[arXiv:1507.08131 [gr-qc]].

\bibitem{Atamurotov:2015xfa}
F.~Atamurotov, S.~G.~Ghosh and B.~Ahmedov,
Eur. Phys. J. C \textbf{76}, no.5, 273 (2016)
doi:10.1140/epjc/s10052-016-4122-9
[arXiv:1506.03690 [gr-qc]].

\bibitem{Cunha:2018acu}
P.~V.~P.~Cunha and C.~A.~R.~Herdeiro,
Gen. Rel. Grav. \textbf{50}, no.4, 42 (2018)
doi:10.1007/s10714-018-2361-9
[arXiv:1801.00860 [gr-qc]].

\bibitem{Cunha:2016wzk}
P.~V.~P.~Cunha, C.~A.~R.~Herdeiro, B.~Kleihaus, J.~Kunz and E.~Radu,
Phys. Lett. B \textbf{768}, 373-379 (2017)
doi:10.1016/j.physletb.2017.03.020
[arXiv:1701.00079 [gr-qc]].

\bibitem{Herdeiro:2021lwl}
C.~A.~R.~Herdeiro, A.~M.~Pombo, E.~Radu, P.~V.~P.~Cunha and N.~Sanchis-Gual,
JCAP \textbf{04}, 051 (2021)
doi:10.1088/1475-7516/2021/04/051
[arXiv:2102.01703 [gr-qc]].

\bibitem{Kuang:2022xjp}
X.~M.~Kuang and A.~\"Ovg\"un,
Annals Phys. \textbf{447}, 169147 (2022)
doi:10.1016/j.aop.2022.169147
[arXiv:2205.11003 [gr-qc]].

\bibitem{Kuang:2022ojj}
X.~M.~Kuang, Z.~Y.~Tang, B.~Wang and A.~Wang,
Phys. Rev. D \textbf{106}, no.6, 064012 (2022)
doi:10.1103/PhysRevD.106.064012
[arXiv:2206.05878 [gr-qc]].

\bibitem{Meng:2022kjs}
Y.~Meng, X.~M.~Kuang and Z.~Y.~Tang,
Phys. Rev. D \textbf{106}, no.6, 064006 (2022)
doi:10.1103/PhysRevD.106.064006
[arXiv:2204.00897 [gr-qc]].

\bibitem{Lima:2021las}
H.~C.~D.~Lima, Junior., L.~C.~B.~Crispino, P.~V.~P.~Cunha and C.~A.~R.~Herdeiro,
Phys. Rev. D \textbf{103}, no.8, 084040 (2021)
doi:10.1103/PhysRevD.103.084040
[arXiv:2102.07034 [gr-qc]].

\bibitem{Pantig:2022toh}
R.~C.~Pantig and A.~\"Ovg\"un,
Eur. Phys. J. C \textbf{82}, no.5, 391 (2022)
doi:10.1140/epjc/s10052-022-10319-8
[arXiv:2201.03365 [gr-qc]].

\bibitem{Lobos:2022jsz}
N.~J.~L.~S.~Lobos and R.~C.~Pantig,
MDPI Physics \textbf{4}, no.4, 1318-1330 (2022)
doi:10.3390/physics4040084
[arXiv:2208.00618 [gr-qc]].

\bibitem{Rayimbaev:2022hca}
J.~Rayimbaev, R.~C.~Pantig, A.~\"Ovg\"un, A.~Abdujabbarov and D.~Demir,
Annals Phys. \textbf{454}, 169335 (2023)
doi:10.1016/j.aop.2023.169335
[arXiv:2206.06599 [gr-qc]].

\bibitem{Pantig:2022ely}
R.~C.~Pantig and A.~\"Ovg\"un,
Annals Phys. \textbf{448}, 169197 (2023)
doi:10.1016/j.aop.2022.169197
[arXiv:2206.02161 [gr-qc]].

\bibitem{Pantig:2022whj}
R.~C.~Pantig and A.~\"Ovg\"un,
JCAP \textbf{08}, no.08, 056 (2022)
doi:10.1088/1475-7516/2022/08/056
[arXiv:2202.07404 [astro-ph.GA]].

\bibitem{Pantig:2022sjb}
R.~C.~Pantig and A.~\"Ovg\"un,
Fortsch. Phys. \textbf{71}, no.1, 2200164 (2023)
doi:10.1002/prop.202200164
[arXiv:2210.00523 [gr-qc]].

\bibitem{Contreras:2019nih}
E.~Contreras, J.~M.~Ramirez-Velasquez, \'A.~Rinc\'on, G.~Panotopoulos and P.~Bargue\~no,
Eur. Phys. J. C \textbf{79}, no.9, 802 (2019)
doi:10.1140/epjc/s10052-019-7309-z
[arXiv:1905.11443 [gr-qc]].

\bibitem{Khodadi:2022ulo}
M.~Khodadi,
Nucl. Phys. B \textbf{985}, 116014 (2022)
doi:10.1016/j.nuclphysb.2022.116014
[arXiv:2211.00300 [gr-qc]].

\bibitem{Badia:2022phg}
J.~Bad\'\i{}a and E.~F.~Eiroa,
[arXiv:2210.03081 [gr-qc]].

\bibitem{Pantig:2022gih}
R.~C.~Pantig, L.~Mastrototaro, G.~Lambiase and A.~\"Ovg\"un,
Eur. Phys. J. C \textbf{82}, no.12, 1155 (2022)
doi:10.1140/epjc/s10052-022-11125-y
[arXiv:2208.06664 [gr-qc]].

\bibitem{Pantig:2022qak}
R.~C.~Pantig, A.~\"Ovg\"un and D.~Demir,
Eur. Phys. J. C \textbf{83}, no.3, 250 (2023)
doi:10.1140/epjc/s10052-023-11400-6
[arXiv:2208.02969 [gr-qc]].

\bibitem{Atamurotov:2022wsr}
F.~Atamurotov, D.~Ortiqboev, A.~Abdujabbarov and G.~Mustafa,
Eur. Phys. J. C \textbf{82}, no.8, 659 (2022)
doi:10.1140/epjc/s10052-022-10619-z

\bibitem{Contreras:2019cmf}
E.~Contreras, \'A.~Rinc\'on, G.~Panotopoulos, P.~Bargue\~no and B.~Koch,
Phys. Rev. D \textbf{101}, no.6, 064053 (2020)
doi:10.1103/PhysRevD.101.064053
[arXiv:1906.06990 [gr-qc]].

\bibitem{Lambiase:2022ywp}
G.~Lambiase and L.~Mastrototaro,
doi:10.3847/1538-4357/ac7140
[arXiv:2205.09785 [hep-ph]].

\bibitem{Kumar:2020hgm}
R.~Kumar, S.~G.~Ghosh and A.~Wang,
Phys. Rev. D \textbf{101}, no.10, 104001 (2020)
doi:10.1103/PhysRevD.101.104001
[arXiv:2001.00460 [gr-qc]].

\bibitem{Kumar:2018ple}
R.~Kumar and S.~G.~Ghosh,
Astrophys. J. \textbf{892}, 78 (2020)
doi:10.3847/1538-4357/ab77b0
[arXiv:1811.01260 [gr-qc]].

\bibitem{Saghafi:2022pme}
S.~Saghafi and K.~Nozari,
JHAP \textbf{3}, no.1, 31-38 (2022)
doi:10.22128/jhap.2022.515.1019

\bibitem{Guo:2022nto}
W.~D.~Guo, S.~W.~Wei and Y.~X.~Liu,
Eur. Phys. J. C \textbf{83}, no.3, 197 (2023)
doi:10.1140/epjc/s10052-023-11316-1
[arXiv:2203.13477 [gr-qc]].

\bibitem{Cimdiker:2021cpz}
\.I.~\c{C}imdiker, D.~Demir and A.~\"Ovg\"un,
Phys. Dark Univ. \textbf{34}, 100900 (2021)
doi:10.1016/j.dark.2021.100900
[arXiv:2110.11904 [gr-qc]].

\bibitem{Xu:2021xgw}
Z.~Xu and M.~Tang,
Chin. Phys. C \textbf{46}, no.8, 085101 (2022)
doi:10.1088/1674-1137/ac6665
[arXiv:2109.14245 [hep-th]].

\bibitem{Konoplya:2022hbl}
R.~A.~Konoplya and A.~Zhidenko,
Astrophys. J. \textbf{933}, no.2, 166 (2022)
doi:10.3847/1538-4357/ac76bc
[arXiv:2202.02205 [gr-qc]].

\bibitem{Atamurotov:2013sca}
F.~Atamurotov, A.~Abdujabbarov and B.~Ahmedov,
Phys. Rev. D \textbf{88}, no.6, 064004 (2013)
doi:10.1103/PhysRevD.88.064004

\bibitem{Konoplya:2019sns}
R.~A.~Konoplya,
Phys. Lett. B \textbf{795}, 1-6 (2019)
doi:10.1016/j.physletb.2019.05.043
[arXiv:1905.00064 [gr-qc]].

\bibitem{Sokoliuk:2022owk}
O.~Sokoliuk, S.~Praharaj, A.~Baransky and P.~K.~Sahoo,
Astron. Astrophys. \textbf{665}, A139 (2022)
doi:10.1051/0004-6361/202244358
[arXiv:2207.07193 [gr-qc]].

\bibitem{Ovgun:2023ego}
A.~\"Ovg\"un, R.~C.~Pantig and \'A.~Rinc\'on,
Eur. Phys. J. Plus \textbf{138}, no.3, 192 (2023)
doi:10.1140/epjp/s13360-023-03793-w
[arXiv:2303.01696 [gr-qc]].

\bibitem{Pantig:2023yer}
R.~C.~Pantig,
[arXiv:2303.01698 [gr-qc]].

\bibitem{Ovgun:2018tua}
A.~\"Ovg\"un, \.I.~Sakall\i{} and J.~Saavedra,
JCAP \textbf{10}, 041 (2018)
doi:10.1088/1475-7516/2018/10/041
[arXiv:1807.00388 [gr-qc]].

\bibitem{Uniyal:2022vdu}
A.~Uniyal, R.~C.~Pantig and A.~\"Ovg\"un,
Phys. Dark Univ. \textbf{40}, 101178 (2023)
doi:10.1016/j.dark.2023.101178
[arXiv:2205.11072 [gr-qc]].

\bibitem{Vagnozzi:2019apd}
S.~Vagnozzi and L.~Visinelli,
Phys. Rev. D \textbf{100}, no.2, 024020 (2019)
doi:10.1103/PhysRevD.100.024020
[arXiv:1905.12421 [gr-qc]].

\bibitem{Khodadi:2021gbc}
M.~Khodadi, G.~Lambiase and D.~F.~Mota,
JCAP \textbf{09}, 028 (2021)
doi:10.1088/1475-7516/2021/09/028
[arXiv:2107.00834 [gr-qc]].

\bibitem{Roy:2020dyy}
R.~Roy and S.~Chakrabarti,
Phys. Rev. D \textbf{102}, no.2, 024059 (2020)
doi:10.1103/PhysRevD.102.024059
[arXiv:2003.14107 [gr-qc]].

\bibitem{Virbhadra:1999nm}
K.~S.~Virbhadra and G.~F.~R.~Ellis,
Phys. Rev. D \textbf{62}, 084003 (2000)
doi:10.1103/PhysRevD.62.084003
[arXiv:astro-ph/9904193 [astro-ph]].

\bibitem{Virbhadra:2002ju}
K.~S.~Virbhadra and G.~F.~R.~Ellis,
Phys. Rev. D \textbf{65}, 103004 (2002)
doi:10.1103/PhysRevD.65.103004

\bibitem{Adler:2022qtb}
S.~L.~Adler and K.~S.~Virbhadra,
Gen. Rel. Grav. \textbf{54}, no.8, 93 (2022)
doi:10.1007/s10714-022-02976-7
[arXiv:2205.04628 [gr-qc]].

\bibitem{Bozza:2001xd}
V.~Bozza, S.~Capozziello, G.~Iovane and G.~Scarpetta,
Gen. Rel. Grav. \textbf{33}, 1535-1548 (2001)
doi:10.1023/A:1012292927358
[arXiv:gr-qc/0102068 [gr-qc]].

\bibitem{Bozza:2002zj}
V.~Bozza,
Phys. Rev. D \textbf{66}, 103001 (2002)
doi:10.1103/PhysRevD.66.103001
[arXiv:gr-qc/0208075 [gr-qc]].

\bibitem{Perlick:2003vg}
V.~Perlick,
Phys. Rev. D \textbf{69}, 064017 (2004)
doi:10.1103/PhysRevD.69.064017
[arXiv:gr-qc/0307072 [gr-qc]].

\bibitem{Virbhadra:2022ybp}
K.~S.~Virbhadra,
[arXiv:2204.01792 [gr-qc]].

\bibitem{Virbhadra:2022iiy}
K.~S.~Virbhadra,
Phys. Rev. D \textbf{106}, no.6, 064038 (2022)
doi:10.1103/PhysRevD.106.064038
[arXiv:2204.01879 [gr-qc]].

\bibitem{Gibbons:2008rj}
G.~W.~Gibbons and M.~C.~Werner,
Class. Quant. Grav. \textbf{25}, 235009 (2008)
doi:10.1088/0264-9381/25/23/235009
[arXiv:0807.0854 [gr-qc]].

\bibitem{Werner:2012rc}
M.~C.~Werner,
Gen. Rel. Grav. \textbf{44}, 3047-3057 (2012)
doi:10.1007/s10714-012-1458-9
[arXiv:1205.3876 [gr-qc]].

\bibitem{Ishihara:2016vdc}
A.~Ishihara, Y.~Suzuki, T.~Ono, T.~Kitamura and H.~Asada,
Phys. Rev. D \textbf{94}, no.8, 084015 (2016)
doi:10.1103/PhysRevD.94.084015
[arXiv:1604.08308 [gr-qc]].

\bibitem{Takizawa:2020egm}
K.~Takizawa, T.~Ono and H.~Asada,
Phys. Rev. D \textbf{101}, no.10, 104032 (2020)
doi:10.1103/PhysRevD.101.104032
[arXiv:2001.03290 [gr-qc]].

\bibitem{Ono:2019hkw}
T.~Ono and H.~Asada,
Universe \textbf{5}, no.11, 218 (2019)
doi:10.3390/universe5110218
[arXiv:1906.02414 [gr-qc]].

\bibitem{Ono:2017pie}
T.~Ono, A.~Ishihara and H.~Asada,
Phys. Rev. D \textbf{96}, no.10, 104037 (2017)
doi:10.1103/PhysRevD.96.104037
[arXiv:1704.05615 [gr-qc]].

\bibitem{Asada:2017vxl}
H.~Asada,
Mod. Phys. Lett. A \textbf{32}, no.34, 1730031 (2017)
doi:10.1142/S0217732317300312
[arXiv:1711.01730 [gr-qc]].

\bibitem{Ovgun:2018fnk}
A.~\"Ovg\"un,
Phys. Rev. D \textbf{98}, no.4, 044033 (2018)
doi:10.1103/PhysRevD.98.044033
[arXiv:1805.06296 [gr-qc]].

\bibitem{Ovgun:2019wej}
A.~\"Ovg\"un,
Phys. Rev. D \textbf{99}, no.10, 104075 (2019)
doi:10.1103/PhysRevD.99.104075
[arXiv:1902.04411 [gr-qc]].

\bibitem{Ovgun:2018oxk}
A.~\"Ovg\"un,
Universe \textbf{5}, no.5, 115 (2019)
doi:10.3390/universe5050115
[arXiv:1806.05549 [physics.gen-ph]].

\bibitem{Ovgun:2020gjz}
A.~\"Ovg\"un and \.I.~Sakall\i{},
Class. Quant. Grav. \textbf{37}, no.22, 225003 (2020)
doi:10.1088/1361-6382/abb579
[arXiv:2005.00982 [gr-qc]].

\bibitem{Ovgun:2018prw}
A.~\"Ovg\"un, G.~Gyulchev and K.~Jusufi,
Annals Phys. \textbf{406}, 152-172 (2019)
doi:10.1016/j.aop.2019.04.007
[arXiv:1806.03719 [gr-qc]].

\bibitem{Ovgun:2018ran}
A.~Ovg\"un, K.~Jusufi and I.~Sakalli,
Annals Phys. \textbf{399}, 193-203 (2018)
doi:10.1016/j.aop.2018.10.012
[arXiv:1805.09431 [gr-qc]].

\bibitem{Javed:2019ynm}
W.~Javed, R.~Babar and A.~\"Ovg\"un,
Phys. Rev. D \textbf{100}, no.10, 104032 (2019)
doi:10.1103/PhysRevD.100.104032
[arXiv:1910.11697 [gr-qc]].

\bibitem{Li:2020dln}
Z.~Li and A.~\"Ovg\"un,
Phys. Rev. D \textbf{101}, no.2, 024040 (2020)
doi:10.1103/PhysRevD.101.024040
[arXiv:2001.02074 [gr-qc]].

\bibitem{Li:2020wvn}
Z.~Li, G.~Zhang and A.~\"Ovg\"un,
Phys. Rev. D \textbf{101}, no.12, 124058 (2020)
doi:10.1103/PhysRevD.101.124058
[arXiv:2006.13047 [gr-qc]].

\bibitem{Belhaj:2022vte}
A.~Belhaj, H.~Belmahi, M.~Benali and H.~Moumni El,
Chin. J. Phys. \textbf{80}, 229-238 (2022)
doi:10.1016/j.cjph.2022.04.013
[arXiv:2204.10150 [gr-qc]].

\bibitem{Javed:2023iih}
W.~Javed, M.~Atique, R.~C.~Pantig and A.~\"Ovg\"un,
Symmetry \textbf{15}, no.1, 148 (2023)
doi:10.3390/sym15010148
[arXiv:2301.01855 [gr-qc]].

\bibitem{Javed:2020lsg}
W.~Javed, A.~Hamza and A.~\"Ovg\"un,
Phys. Rev. D \textbf{101}, no.10, 103521 (2020)
doi:10.20944/preprints201911.0142.v1
[arXiv:2005.09464 [gr-qc]].

\bibitem{Javed:2022psa}
W.~Javed, M.~Atique, R.~C.~Pantig and A.~\"Ovg\"un,
Int. J. Geom. Meth. Mod. Phys. \textbf{20}, no.03, 2350040 (2023)
doi:10.1142/S0219887823500408

\bibitem{Javed:2022fsn}
W.~Javed, S.~Riaz, R.~C.~Pantig and A.~\"Ovg\"un,
Eur. Phys. J. C \textbf{82}, no.11, 1057 (2022)
doi:10.1140/epjc/s10052-022-11030-4
[arXiv:2212.00804 [gr-qc]].

\bibitem{Javed:2022gtz}
W.~Javed, H.~Irshad, R.~C.~Pantig and A.~\"Ovg\"un,
Universe \textbf{8}, no.11, 599 (2022)
doi:10.3390/universe8110599
[arXiv:2211.07009 [gr-qc]].

\bibitem{Ovgun:2020yuv}
A.~\"Ovg\"un,
Turk. J. Phys. \textbf{44}, no.5, 465-471 (2020)
doi:10.20944/preprints202008.0512.v1
[arXiv:2011.04423 [gr-qc]].

\bibitem{Javed:2019jag}
W.~Javed, J.~Abbas and A.~\"Ovg\"un,
Annals Phys. \textbf{418}, 168183 (2020)
doi:10.20944/preprints201906.0124.v1
[arXiv:2007.16027 [gr-qc]].

\bibitem{Javed:2019rrg}
W.~Javed, j.~Abbas and A.~\"Ovg\"un,
Phys. Rev. D \textbf{100}, no.4, 044052 (2019)
doi:10.20944/preprints201906.0101.v1
[arXiv:1908.05241 [gr-qc]].

\bibitem{Kumaran:2019qqp}
Y.~Kumaran and A.~\"Ovg\"un,
Chin. Phys. C \textbf{44}, no.2, 025101 (2020)
doi:10.1088/1674-1137/44/2/025101
[arXiv:1905.11710 [gr-qc]].

\bibitem{Jusufi:2017mav}
K.~Jusufi and A.~\"Ovg\"un,
Phys. Rev. D \textbf{97}, no.2, 024042 (2018)
doi:10.1103/PhysRevD.97.024042
[arXiv:1708.06725 [gr-qc]].

\bibitem{Javed:2019qyg}
W.~Javed, R.~Babar and A.~\"Ovg\"un,
Phys. Rev. D \textbf{99}, no.8, 084012 (2019)
doi:10.1103/PhysRevD.99.084012
[arXiv:1903.11657 [gr-qc]].

\bibitem{Parikh:1999mf}
M.~K.~Parikh and F.~Wilczek,
Phys. Rev. Lett. \textbf{85}, 5042-5045 (2000)
doi:10.1103/PhysRevLett.85.5042
[arXiv:hep-th/9907001 [hep-th]].

\bibitem{Singleton:2011vh}
D.~Singleton and S.~Wilburn,
Phys. Rev. Lett. \textbf{107}, 081102 (2011)
doi:10.1103/PhysRevLett.107.081102
[arXiv:1102.5564 [gr-qc]].

\bibitem{Akhmedova:2008dz}
V.~Akhmedova, T.~Pilling, A.~de Gill and D.~Singleton,
Phys. Lett. B \textbf{666}, 269-271 (2008)
doi:10.1016/j.physletb.2008.07.017
[arXiv:0804.2289 [hep-th]].

\bibitem{Maldacena:1996ix}
J.~M.~Maldacena and A.~Strominger,
Phys. Rev. D \textbf{55}, 861-870 (1997)
doi:10.1103/PhysRevD.55.861
[arXiv:hep-th/9609026 [hep-th]].

\bibitem{Cvetic:1997uw}
M.~Cvetic and F.~Larsen,
Phys. Rev. D \textbf{56}, 4994-5007 (1997)
doi:10.1103/PhysRevD.56.4994
[arXiv:hep-th/9705192 [hep-th]].

\bibitem{Harmark:2007jy}
T.~Harmark, J.~Natario and R.~Schiappa,
Adv. Theor. Math. Phys. \textbf{14}, no.3, 727-794 (2010)
doi:10.4310/ATMP.2010.v14.n3.a1
[arXiv:0708.0017 [hep-th]].

\bibitem{Gubser:1996zp}
S.~S.~Gubser and I.~R.~Klebanov,
Phys. Rev. Lett. \textbf{77}, 4491-4494 (1996)
doi:10.1103/PhysRevLett.77.4491
[arXiv:hep-th/9609076 [hep-th]].

\bibitem{Zhang:2020qam}
C.~Y.~Zhang, P.~C.~Li and M.~Guo,
Eur. Phys. J. C \textbf{80}, no.9, 874 (2020)
doi:10.1140/epjc/s10052-020-08448-z
[arXiv:2003.13068 [hep-th]].

\bibitem{Fernando:2004ay}
S.~Fernando,
``Greybody factors of charged dilaton black holes in 2 + 1 dimensions,''
Gen. Rel. Grav. \textbf{37}, 461-481 (2005)
doi:10.1007/s10714-005-0035-x
[arXiv:hep-th/0407163 [hep-th]].

\bibitem{Kanti:2014dxa}
P.~Kanti, T.~Pappas and N.~Pappas,
Phys. Rev. D \textbf{90}, no.12, 124077 (2014)
doi:10.1103/PhysRevD.90.124077
[arXiv:1409.8664 [hep-th]].

\bibitem{Ovgun:2018gwt}
A.~Ovg\"un and K.~Jusufi,
Annals Phys. \textbf{395}, 138-151 (2018)
doi:10.1016/j.aop.2018.05.013
[arXiv:1801.02555 [gr-qc]].

\bibitem{Konoplya:2003ii}
R.~A.~Konoplya,
Phys. Rev. D \textbf{68}, 024018 (2003)
doi:10.1103/PhysRevD.68.024018
[arXiv:gr-qc/0303052 [gr-qc]].

\bibitem{Berti:2009kk}
E.~Berti, V.~Cardoso and A.~O.~Starinets,
Class. Quant. Grav. \textbf{26}, 163001 (2009)
doi:10.1088/0264-9381/26/16/163001
[arXiv:0905.2975 [gr-qc]].

\bibitem{Konoplya:2011qq}
R.~A.~Konoplya and A.~Zhidenko,
Rev. Mod. Phys. \textbf{83}, 793-836 (2011)
doi:10.1103/RevModPhys.83.793
[arXiv:1102.4014 [gr-qc]].



\bibitem{Zhidenko:2005mv}
A.~Zhidenko,
Class. Quant. Grav. \textbf{23}, 3155-3164 (2006)
doi:10.1088/0264-9381/23/9/024
[arXiv:gr-qc/0510039 [gr-qc]].

\bibitem{Daghigh:2008jz}
R.~G.~Daghigh and M.~D.~Green,
Class. Quant. Grav. \textbf{26}, 125017 (2009)
doi:10.1088/0264-9381/26/12/125017
[arXiv:0808.1596 [gr-qc]].

\bibitem{Daghigh:2011ty}
R.~G.~Daghigh and M.~D.~Green,
Phys. Rev. D \textbf{85}, 127501 (2012)
doi:10.1103/PhysRevD.85.127501
[arXiv:1112.5397 [gr-qc]].

\bibitem{Okyay:2021nnh}
M.~Okyay and A.~\"Ovg\"un,
JCAP \textbf{01}, no.01, 009 (2022)
doi:10.1088/1475-7516/2022/01/009
[arXiv:2108.07766 [gr-qc]].

\bibitem{Panotopoulos:2018pvu}
G.~Panotopoulos and A.~Rinc\'on,
Phys. Rev. D \textbf{97}, no.8, 085014 (2018)
doi:10.1103/PhysRevD.97.085014
[arXiv:1804.04684 [hep-th]].

\bibitem{Panotopoulos:2016wuu}
G.~Panotopoulos and \'A.~Rinc\'on,
Phys. Lett. B \textbf{772}, 523-528 (2017)
doi:10.1016/j.physletb.2017.07.014
[arXiv:1611.06233 [hep-th]].

\bibitem{Rincon:2018ktz}
\'A.~Rinc\'on and G.~Panotopoulos,
Eur. Phys. J. C \textbf{78}, no.10, 858 (2018)
doi:10.1140/epjc/s10052-018-6352-5
[arXiv:1810.08822 [gr-qc]].

\bibitem{Panotopoulos:2017yoe}
G.~Panotopoulos and \'A.~Rinc\'on,
Phys. Rev. D \textbf{96}, no.2, 025009 (2017)
doi:10.1103/PhysRevD.96.025009
[arXiv:1706.07455 [hep-th]].

\bibitem{Ahmed:2016lou}
J.~Ahmed and K.~Saifullah,
Eur. Phys. J. C \textbf{78}, no.4, 316 (2018)
doi:10.1140/epjc/s10052-018-5800-6
[arXiv:1610.06104 [gr-qc]].

\bibitem{Al-Badawi:2022aby}
A.~Al-Badawi, S.~Kanzi and \.I.~Sakall\i{},
Annals Phys. \textbf{452}, 169294 (2023)
doi:10.1016/j.aop.2023.169294
[arXiv:2203.04140 [hep-th]].

\bibitem{Al-Badawi:2021wdm}
A.~Al-Badawi, S.~Kanzi and \.I.~Sakall\i{},
Eur. Phys. J. Plus \textbf{137}, no.1, 94 (2022)
doi:10.1140/epjp/s13360-021-02227-9
[arXiv:2111.15005 [gr-qc]].

\bibitem{Javed:2021ymu}
W.~Javed, I.~Hussain and A.~\"Ovg\"un,
Eur. Phys. J. Plus \textbf{137}, no.1, 148 (2022)
doi:10.1140/epjp/s13360-022-02374-7
[arXiv:2201.09879 [gr-qc]].

\bibitem{Javed:2022kzf}
W.~Javed, M.~Aqib and A.~\"Ovg\"un,
Phys. Lett. B \textbf{829}, 137114 (2022)
doi:10.1016/j.physletb.2022.137114
[arXiv:2204.07864 [gr-qc]].

\bibitem{Javed:2022rrs}
W.~Javed, S.~Riaz and A.~\"Ovg\"un,
Universe \textbf{8}, no.5, 262 (2022)
doi:10.3390/universe8050262
[arXiv:2205.02229 [gr-qc]].

\bibitem{Mangut:2023oxa}
M.~Mangut, H.~G\"ursel, S.~Kanzi and \.I.~Sakall\i{},
Universe \textbf{9}, no.5, 225 (2023)
doi:10.3390/universe9050225
[arXiv:2305.10815 [gr-qc]].

\bibitem{Sakalli:2023pgn}
\.I.~Sakall\i{} and E.~Y\"or\"uk,
Phys. Scripta \textbf{98}, no.12, 125307 (2023)
doi:10.1088/1402-4896/ad09a1

\bibitem{Lee:1998pd}
H.~W.~Lee and Y.~S.~Myung,
Phys. Rev. D \textbf{58}, 104013 (1998)
doi:10.1103/PhysRevD.58.104013
[arXiv:hep-th/9804095 [hep-th]].

\bibitem{Pappas:2016ovo}
T.~Pappas, P.~Kanti and N.~Pappas,
Phys. Rev. D \textbf{94}, no.2, 024035 (2016)
doi:10.1103/PhysRevD.94.024035
[arXiv:1604.08617 [hep-th]].

\bibitem{Crispino:2013pya}
L.~C.~B.~Crispino, A.~Higuchi, E.~S.~Oliveira and J.~V.~Rocha,
Phys. Rev. D \textbf{87}, 104034 (2013)
doi:10.1103/PhysRevD.87.104034
[arXiv:1304.0467 [gr-qc]].

\bibitem{Chen:2010ru}
S.~Chen and J.~Jing,
Phys. Lett. B \textbf{691}, 254-260 (2010)
doi:10.1016/j.physletb.2010.06.041
[arXiv:1005.5601 [gr-qc]].

\bibitem{Devi:2020uac}
S.~Devi, R.~Roy and S.~Chakrabarti,
Eur. Phys. J. C \textbf{80}, no.8, 760 (2020)
doi:10.1140/epjc/s10052-020-8311-1
[arXiv:2004.14935 [gr-qc]].

\bibitem{Gogoi:2022wyv}
D.~J.~Gogoi and U.~D.~Goswami,
JCAP \textbf{06}, no.06, 029 (2022)
doi:10.1088/1475-7516/2022/06/029
[arXiv:2203.07594 [gr-qc]].

\bibitem{Kanti:2009sn}
P.~Kanti, H.~Kodama, R.~A.~Konoplya, N.~Pappas and A.~Zhidenko,
Phys. Rev. D \textbf{80}, 084016 (2009)
doi:10.1103/PhysRevD.80.084016
[arXiv:0906.3845 [hep-th]].

\bibitem{Visser:1998ke}
M.~Visser,
Phys. Rev. A \textbf{59}, 427-438 (1999)
doi:10.1103/PhysRevA.59.427
[arXiv:quant-ph/9901030 [quant-ph]].


\bibitem{Boonserm:2008zg}
P.~Boonserm and M.~Visser,
Phys. Rev. D \textbf{78}, 101502 (2008)
doi:10.1103/PhysRevD.78.101502
[arXiv:0806.2209 [gr-qc]].

\bibitem{Boonserm:2017qcq}
P.~Boonserm, T.~Ngampitipan and P.~Wongjun,
Eur. Phys. J. C \textbf{78}, no.6, 492 (2018)
doi:10.1140/epjc/s10052-018-5975-x
[arXiv:1705.03278 [gr-qc]].

\bibitem{Boonserm:2019mon}
P.~Boonserm, T.~Ngampitipan and P.~Wongjun,
Eur. Phys. J. C \textbf{79}, no.4, 330 (2019)
doi:10.1140/epjc/s10052-019-6827-z
[arXiv:1902.05215 [gr-qc]].

\bibitem{Liu:2022ygf}
D.~Liu, Y.~Yang, A.~\"Ovg\"un, Z.~W.~Long and Z.~Xu,
Eur. Phys. J. C \textbf{83}, no.7, 565 (2023)
doi:10.1140/epjc/s10052-023-11739-w
[arXiv:2204.11563 [gr-qc]].

\bibitem{Yang:2022ifo}
Y.~Yang, D.~Liu, A.~\"Ovg\"un, Z.~W.~Long and Z.~Xu,
Phys. Rev. D \textbf{107}, no.6, 064042 (2023)
doi:10.1103/PhysRevD.107.064042
[arXiv:2203.11551 [gr-qc]].

\bibitem{Gray:2015xig}
F.~Gray and M.~Visser,
Universe \textbf{4}, no.9, 93 (2018)
doi:10.3390/universe4090093
[arXiv:1512.05018 [gr-qc]].

\bibitem{Boonserm:2014fja}
P.~Boonserm, A.~Chatrabhuti, T.~Ngampitipan and M.~Visser,
J. Math. Phys. \textbf{55}, 112502 (2014)
doi:10.1063/1.4901127
[arXiv:1405.5678 [gr-qc]].

\bibitem{Boonserm:2014rma}
P.~Boonserm, T.~Ngampitipan and M.~Visser,
JHEP \textbf{03}, 113 (2014)
doi:10.1007/JHEP03(2014)113
[arXiv:1401.0568 [gr-qc]].

\bibitem{Boonserm:2013dua}
P.~Boonserm, T.~Ngampitipan and M.~Visser,
Phys. Rev. D \textbf{88}, 041502 (2013)
doi:10.1103/PhysRevD.88.041502
[arXiv:1305.1416 [gr-qc]].

\bibitem{Ngampitipan:2012dq}
T.~Ngampitipan and P.~Boonserm,
Int. J. Mod. Phys. D \textbf{22}, 1350058 (2013)
doi:10.1142/S0218271813500582
[arXiv:1211.4070 [math-ph]].

\bibitem{Boonserm:2009zba}
P.~Boonserm,
[arXiv:0907.0045 [math-ph]].

\bibitem{Boonserm:2009mi}
P.~Boonserm and M.~Visser,
Annals Phys. \textbf{325}, 1328-1339 (2010)
doi:10.1016/j.aop.2010.02.005
[arXiv:0901.0944 [math-ph]].

\bibitem{Lambiase:2023zeo}
G.~Lambiase, L.~Mastrototaro, R.~C.~Pantig and A.~Ovgun,
``Probing Schwarzschild-like black holes in metric-affine bumblebee gravity with accretion disk, deflection angle, greybody bounds, and neutrino propagation,''
JCAP \textbf{12}, 026 (2023)
doi:10.1088/1475-7516/2023/12/026
[arXiv:2309.13594 [gr-qc]].

\bibitem{Gogoi:2023fow}
D.~J.~Gogoi, A.~\"Ovg\"un and D.~Demir,
``Quasinormal modes and greybody factors of symmergent black hole,''
Phys. Dark Univ. \textbf{42}, 101314 (2023)
doi:10.1016/j.dark.2023.101314
[arXiv:2306.09231 [gr-qc]].



\bibitem{Padmanabhan:2001ev}
T.~Padmanabhan,
Class. Quant. Grav. \textbf{19}, 3551-3566 (2002)
doi:10.1088/0264-9381/19/13/312
[arXiv:gr-qc/0110046 [gr-qc]].

\bibitem{Padmanabhan:1998yy}
T.~Padmanabhan,
[arXiv:hep-th/9812018 [hep-th]].

\bibitem{Reuter:2003ca}
M.~Reuter and H.~Weyer,
Phys. Rev. D \textbf{69}, 104022 (2004)
doi:10.1103/PhysRevD.69.104022
[arXiv:hep-th/0311196 [hep-th]].

\bibitem{Rincon:2022hpy}
A.~Rincon, B.~Koch, C.~Laporte, F.~Canales and N.~Cruz,
Eur. Phys. J. C \textbf{83}, no.2, 105 (2023)
doi:10.1140/epjc/s10052-023-11169-8
[arXiv:2212.13623 [gr-qc]].

\bibitem{Bonanno:2000ep}
A.~Bonanno and M.~Reuter,
Phys. Rev. D \textbf{62}, 043008 (2000)
doi:10.1103/PhysRevD.62.043008
[arXiv:hep-th/0002196 [hep-th]].

\bibitem{Koch:2014cqa}
B.~Koch and F.~Saueressig,
Int. J. Mod. Phys. A \textbf{29}, no.8, 1430011 (2014)
doi:10.1142/S0217751X14300117
[arXiv:1401.4452 [hep-th]].

\bibitem{Cai:2010zh}
Y.~F.~Cai and D.~A.~Easson,
JCAP \textbf{09}, 002 (2010)
doi:10.1088/1475-7516/2010/09/002
[arXiv:1007.1317 [hep-th]].

\bibitem{Gonzalez:2015upa}
C.~Gonz\'alez and B.~Koch,
Int. J. Mod. Phys. A \textbf{31}, no.26, 1650141 (2016)
doi:10.1142/S0217751X16501414
[arXiv:1508.01502 [hep-th]].

\bibitem{Platania:2023srt}
A.~Platania,
[arXiv:2302.04272 [gr-qc]].

\bibitem{Ishibashi:2021kmf}
A.~Ishibashi, N.~Ohta and D.~Yamaguchi,
Phys. Rev. D \textbf{104}, no.6, 066016 (2021)
doi:10.1103/PhysRevD.104.066016
[arXiv:2106.05015 [hep-th]].

\bibitem{Contreras:2017eza}
E.~Contreras, \'A.~Rinc\'on, B.~Koch and P.~Bargue\~no,
Int. J. Mod. Phys. D \textbf{27}, no.03, 1850032 (2017)
doi:10.1142/S0218271818500323
[arXiv:1711.08400 [gr-qc]].

\bibitem{Koch:2014joa}
B.~Koch, P.~Rioseco and C.~Contreras,
Phys. Rev. D \textbf{91}, no.2, 025009 (2015)
doi:10.1103/PhysRevD.91.025009
[arXiv:1409.4443 [hep-th]].

\bibitem{Rincon:2017ayr}
A.~Rincon and B.~Koch,
J. Phys. Conf. Ser. \textbf{1043}, no.1, 012015 (2018)
doi:10.1088/1742-6596/1043/1/012015
[arXiv:1705.02729 [hep-th]].

\bibitem{Lambiase:2022xde}
G.~Lambiase and F.~Scardigli,
Phys. Rev. D \textbf{105} (2022) no.12, 124054
[arXiv:2204.07416 [hep-th]].

\bibitem{Scardigli:2022jtt}
F.~Scardigli and G.~Lambiase,
Phys. Rev. D \textbf{107} (2023) no.10, 104001
[arXiv:2205.07088 [gr-qc]].

\bibitem{Lambiase:2023hng}
G.~Lambiase, R.~C.~Pantig, D.~J.~Gogoi and A.~\"Ovg\"un,
``Investigating the connection between generalized uncertainty principle and asymptotically safe gravity in black hole signatures through shadow and quasinormal modes,''
Eur. Phys. J. C \textbf{83}, no.7, 679 (2023)
[arXiv:2304.00183 [gr-qc]].

\bibitem{Rincon:2017goj}
\'A.~Rinc\'on, E.~Contreras, P.~Bargue\~no, B.~Koch, G.~Panotopoulos and A.~Hern\'andez-Arboleda,
Eur. Phys. J. C \textbf{77}, no.7, 494 (2017)
doi:10.1140/epjc/s10052-017-5045-9
[arXiv:1704.04845 [hep-th]].

\bibitem{Rincon:2018sgd}
\'A.~Rinc\'on and G.~Panotopoulos,
Phys. Rev. D \textbf{97}, no.2, 024027 (2018)
doi:10.1103/PhysRevD.97.024027
[arXiv:1801.03248 [hep-th]].

\bibitem{Rincon:2018lyd}
\'A.~Rinc\'on and B.~Koch,
Eur. Phys. J. C \textbf{78}, no.12, 1022 (2018)
doi:10.1140/epjc/s10052-018-6488-3
[arXiv:1806.03024 [hep-th]].

\bibitem{Rincon:2019zxk}
\'A.~Rinc\'on and J.~R.~Villanueva,
Class. Quant. Grav. \textbf{37}, no.17, 175003 (2020)
doi:10.1088/1361-6382/aba17f
[arXiv:1902.03704 [gr-qc]].

\bibitem{Fathi:2019jid}
M.~Fathi, \'A.~Rinc\'on and J.~R.~Villanueva,
Class. Quant. Grav. \textbf{37}, no.7, 075004 (2020)
doi:10.1088/1361-6382/ab6f7c
[arXiv:1903.09037 [gr-qc]].

\bibitem{Contreras:2018gpl}
E.~Contreras and P.~Bargue\~no,
Mod. Phys. Lett. A \textbf{33}, no.32, 1850184 (2018)
doi:10.1142/S0217732318501845
[arXiv:1809.00785 [gr-qc]].

\bibitem{Rincon:2020cpz}
A.~Rinc\'on and G.~Panotopoulos,
Phys. Dark Univ. \textbf{30}, 100725 (2020)
doi:10.1016/j.dark.2020.100725
[arXiv:2009.14678 [gr-qc]].

\bibitem{Rincon:2019cix}
\'A.~Rinc\'on, E.~Contreras, P.~Bargue\~no and B.~Koch,
Eur. Phys. J. Plus \textbf{134}, no.11, 557 (2019)
doi:10.1140/epjp/i2019-13081-5
[arXiv:1901.03650 [gr-qc]].

\bibitem{Sendra:2018vux}
C.~M.~Sendra,
Gen. Rel. Grav. \textbf{51}, no.7, 83 (2019)
doi:10.1007/s10714-019-2571-9
[arXiv:1807.07038 [gr-qc]].

\bibitem{Panotopoulos:2021tkk}
G.~Panotopoulos, A.~Rinc\'on and I.~Lopes,
Phys. Rev. D \textbf{103}, 104040 (2021)
doi:10.1103/PhysRevD.103.104040
[arXiv:2104.13611 [gr-qc]].

\bibitem{Contreras:2018swc}
E.~Contreras and P.~Bargue\~no,
Int. J. Mod. Phys. D \textbf{27}, no.09, 1850101 (2018)
doi:10.1142/S0218271818501018
[arXiv:1804.00988 [gr-qc]].

\bibitem{Canales:2018tbn}
F.~Canales, B.~Koch, C.~Laporte and A.~Rincon,
JCAP \textbf{01}, 021 (2020)
doi:10.1088/1475-7516/2020/01/021
[arXiv:1812.10526 [gr-qc]].

\bibitem{Alvarez:2022wef}
P.~D.~Alvarez, B.~Koch, C.~Laporte and A.~Rincon,
[arXiv:2210.11853 [gr-qc]].

\bibitem{Panotopoulos:2021heb}
G.~Panotopoulos and \'A.~Rinc\'on,
Eur. Phys. J. Plus \textbf{136}, no.6, 622 (2021)
doi:10.1140/epjp/s13360-021-01583-w
[arXiv:2105.10803 [gr-qc]].

\bibitem{Alvarez:2022mlf}
P.~D.~Alvarez, B.~Koch, C.~Laporte, F.~Canales and A.~Rincon,
JCAP \textbf{10}, 071 (2022)
doi:10.1088/1475-7516/2022/10/071
[arXiv:2205.05592 [gr-qc]].

\bibitem{Alvarez:2020xmk}
P.~D.~Alvarez, B.~Koch, C.~Laporte and \'A.~Rinc\'on,
JCAP \textbf{06}, 019 (2021)
doi:10.1088/1475-7516/2021/06/019
[arXiv:2009.02311 [gr-qc]].

\bibitem{Bargueno:2021nuc}
P.~Bargue\~no, E.~Contreras and \'A.~Rinc\'on,
Eur. Phys. J. C \textbf{81}, no.5, 477 (2021)
doi:10.1140/epjc/s10052-021-09274-7
[arXiv:2105.10178 [gr-qc]].

\bibitem{Panotopoulos:2021obe}
G.~Panotopoulos, \'A.~Rinc\'on and I.~Lopes,
Eur. Phys. J. C \textbf{81}, no.1, 63 (2021)
doi:10.1140/epjc/s10052-021-08881-8
[arXiv:2101.06649 [gr-qc]].

\bibitem{Panotopoulos:2020zqa}
G.~Panotopoulos, \'A.~Rinc\'on and I.~Lopes,
Eur. Phys. J. C \textbf{80}, no.4, 318 (2020)
doi:10.1140/epjc/s10052-020-7900-3
[arXiv:2004.02627 [gr-qc]].

\bibitem{Biemans:2016rvp}
J.~Biemans, A.~Platania and F.~Saueressig,
Phys. Rev. D \textbf{95} (2017) no.8, 086013
[arXiv:1609.04813 [hep-th]].


\bibitem{EventHorizonTelescope:2021dqv}
P.~Kocherlakota \textit{et al.} [Event Horizon Telescope],
Phys. Rev. D \textbf{103}, no.10, 104047 (2021)
doi:10.1103/PhysRevD.103.104047
[arXiv:2105.09343 [gr-qc]].

\bibitem{Vagnozzi:2022moj}
S.~Vagnozzi, R.~Roy, Y.~D.~Tsai, L.~Visinelli, M.~Afrin, A.~Allahyari, P.~Bambhaniya, D.~Dey, S.~G.~Ghosh and P.~S.~Joshi, \textit{et al.}
[arXiv:2205.07787 [gr-qc]].

\bibitem{Scardigli:2014qka}
F.~Scardigli and R.~Casadio,
Eur. Phys. J. C \textbf{75} (2015) no.9, 425
doi:10.1140/epjc/s10052-015-3635-y
[arXiv:1407.0113 [hep-th]].

\end{thebibliography}
\end{document}